\documentclass[10pt,twocolumn,letterpaper]{article}

\usepackage[pagenumbers]{cvpr}

\usepackage{amsfonts}
\usepackage{mathtools}
\usepackage{array}
\usepackage{textcomp}
\usepackage{verbatim}
\usepackage{capt-of}
\usepackage{siunitx}
\usepackage{multirow}
\usepackage{tabularx}
\usepackage{colortbl}
\usepackage{algorithm}
\usepackage{algpseudocode}
\usepackage{placeins}

\newcolumntype{C}{>{\centering\arraybackslash}X}
\definecolor{rank1}{RGB}{137,190,222}
\definecolor{rank2}{RGB}{177,211,232}
\definecolor{rank3}{RGB}{198,219,239}
\definecolor{rank4}{RGB}{234,242,248}
\newcommand{\rcell}[2]{%
  \ifnum#1>4\relax
    #2%
  \else
    \cellcolor{rank#1}#2%
  \fi
}

\usepackage{stfloats}

\definecolor{cvprblue}{rgb}{0.21,0.49,0.74}
\usepackage[pagebackref,breaklinks,colorlinks,allcolors=cvprblue]{hyperref}

\def\confName{CVPR}
\def\confYear{2027}

\title{Edit2MeshGS: Bridging 3D Scene Editing and Mesh Reconstruction with 2D Gaussian Splatting}

\author{Bingzhi Fan\textsuperscript{1} \quad Wei Jing\textsuperscript{2} \quad
Yaoteng Zhang\textsuperscript{1} \quad Zhitong Xiong\textsuperscript{2} \quad Qi Wang\textsuperscript{2}\\
\normalsize \textsuperscript{1}\,School of Computer Science, Northwestern Polytechnical University, China\\[-1pt]
\normalsize \textsuperscript{2}\,School of Artificial Intelligence, Optics and Electronics (iOPEN), Northwestern Polytechnical University, China}

\hypersetup{
    pdftitle={Edit2MeshGS: Bridging 3D Scene Editing and Mesh Reconstruction with 2D Gaussian Splatting},
    pdfauthor={Bingzhi Fan, Wei Jing, Yaoteng Zhang, Zhitong Xiong, Qi Wang}
}

\makeatletter
\patchcmd{\@maketitle}{\vspace*{24pt}}{\vspace*{9pt}}{}{}
\makeatother

\begin{document}
\twocolumn[{%
\renewcommand\twocolumn[1][]{#1}%
\maketitle
\vspace*{-28.5pt}
\begin{center}
    \centering
    \captionsetup{type=figure}
    \includegraphics[width=0.99\textwidth]{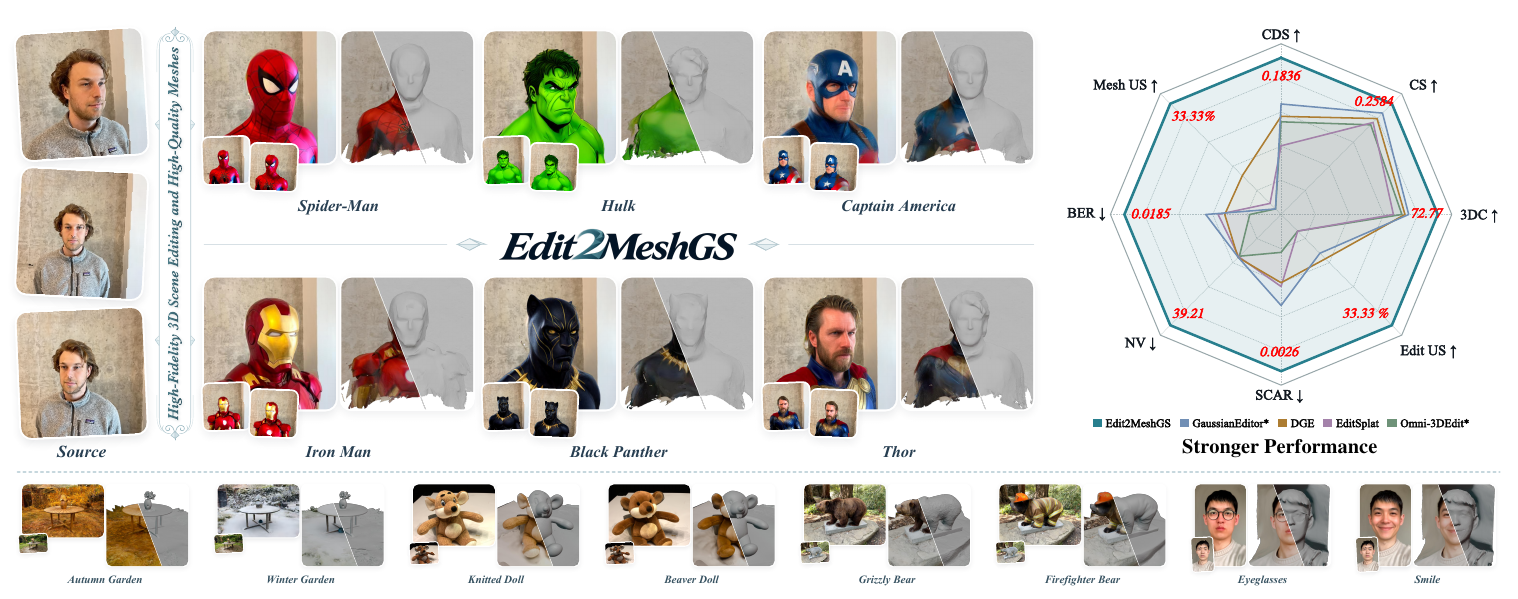}
    \captionof{figure}{Edit2MeshGS achieves both high-fidelity 3D scene editing and reliable mesh reconstruction. We demonstrate diverse edits, including character replacement, seasonal changes, and local attribute editing, with rendered views and corresponding textured and untextured meshes. The radar chart compares editing and mesh quality with representative methods: curves farther from the center indicate better performance, and red numbers show our scores.}
    \label{fig:teaser}
\end{center}%
}]
\begin{abstract}
3D scene editing transforms source scenes into target scenes according to instructions. However, existing Gaussian Splatting-based scene editing methods degrade scene geometry and leave source structures behind, failing to achieve both high-fidelity editing and reliable mesh reconstruction. Therefore, we propose Edit2MeshGS, a framework that achieves both high-fidelity 3D scene editing and reliable mesh reconstruction. The framework uses shared conditioning to guide a diffusion-based editing model toward the intended editing direction, achieving high-fidelity editing. Meanwhile, geometric constraints are introduced during scene optimization to mitigate geometric degradation. Semantic Transition Pruning (STP) is further proposed to model the source-to-target semantic transitions of Gaussians and remove residual source structures, improving Edit2MeshGS's mesh reconstruction quality. Experiments show Edit2MeshGS improves CLIP scores and mesh quality by 41.8\% and 74.0\%, respectively, compared with the second-best method, demonstrating its effectiveness.
\end{abstract}

\section{Introduction}

As the demand for 3D content continues to grow in applications such as digital twins, robotics simulation, and immersive interaction, reusing and adapting existing 3D assets has become increasingly important.
3D scene editing~\cite{haque2023instruct,he2024customize,barda2025instant3dit,chen2024gaussianeditor} enables the modification of existing scenes according to user instructions, faithfully realizing the intended changes while preserving visual realism.
Reconstructing meshes from edited scenes further supports downstream tasks such as modeling, animation, and physical simulation.
Therefore, achieving both high-fidelity editing and reliable mesh reconstruction can extend the applications of edited scenes beyond visual presentation to robotics simulation and interactive environment construction.
However, existing methods primarily focus on either high-fidelity editing or reliable mesh reconstruction, leaving a gap between edited scenes and the meshes required for downstream tasks.

Editing methods~\cite{palandra2024gsedit,xiao2025localized,zhuang2024tip,chen2024dge,wu2024gaussctrl,lee2025editsplat} based on Gaussian Splatting (GS)~\cite{kerbl20233d} have demonstrated significant advantages in 3D scene editing due to their flexible scene representations and high-fidelity editing capabilities.
Nevertheless, \textbf{geometric degradation} during editing and \textbf{the persistence of source structures} after editing hinder reliable mesh reconstruction.
Specifically, geometric degradation primarily arises because existing methods optimize Gaussian scenes using per-view pixel-level supervision without directly constraining the spatial distribution of Gaussians.
This leads to local spatial drift and subsequent surface distortions during mesh reconstruction.
Furthermore, some source Gaussians can satisfy the editing requirements through updates to their colors and opacities alone, leaving the original geometry and introducing redundant structures inconsistent with the intended edits into the reconstructed meshes.
In contrast, mesh-based 3D editing methods~\cite{gao2023textdeformer,michel2022text2mesh} focus on individual objects, making them difficult to extend to high-fidelity editing of complete 3D scenes.
\textit{Therefore, how to achieve high-fidelity editing of complete scenes while ensuring that the edited results can be reconstructed into reliable meshes remains a key challenge.}

In this work, we propose Edit2MeshGS, a framework that achieves both high-fidelity 3D scene editing and reliable mesh reconstruction using 2D Gaussian Splatting (2DGS) \cite{huang20242d}. The framework uses shared conditioning to guide a diffusion-based editing model, aligning editing directions across different views with the intended target and providing high-fidelity editing supervision for Gaussian scene optimization. \textbf{To mitigate geometric degradation}, geometric constraints are introduced to regularize the local surface geometry of Gaussian primitives during editing. \textbf{For residual source structures removal}, Semantic Transition Pruning (STP) is proposed to model source-to-target semantic transitions of Gaussians, identify and remove source structures that persist after editing. Additionally, we introduce Soft Veto Gaussian Tagging (SVGT), which combines multi-view semantic aggregation with per-view verification to enable local editing while protecting the geometry of unedited regions.

Our contributions are summarized as follows:

\begin{itemize}
    \item We propose Edit2MeshGS, a 2DGS-based framework that achieves both high-fidelity 3D scene editing and reliable mesh reconstruction. Experiments demonstrate improvements of 41.8\% in CLIP score and 74.0\% in mesh reconstruction quality over the second-best method.
    \item Semantic Transition Pruning (STP) is proposed to model source-to-target semantic transitions of Gaussians, removing residual source structures to improve mesh reconstruction quality.

    \item Soft Veto Gaussian Tagging (SVGT) is introduced to combine multi-view semantic aggregation with per-view verification, enabling precise local editing while protecting the geometry of unedited regions.
\end{itemize}

\begin{figure*}[t]
    \centering
    \includegraphics[width=0.98\textwidth]{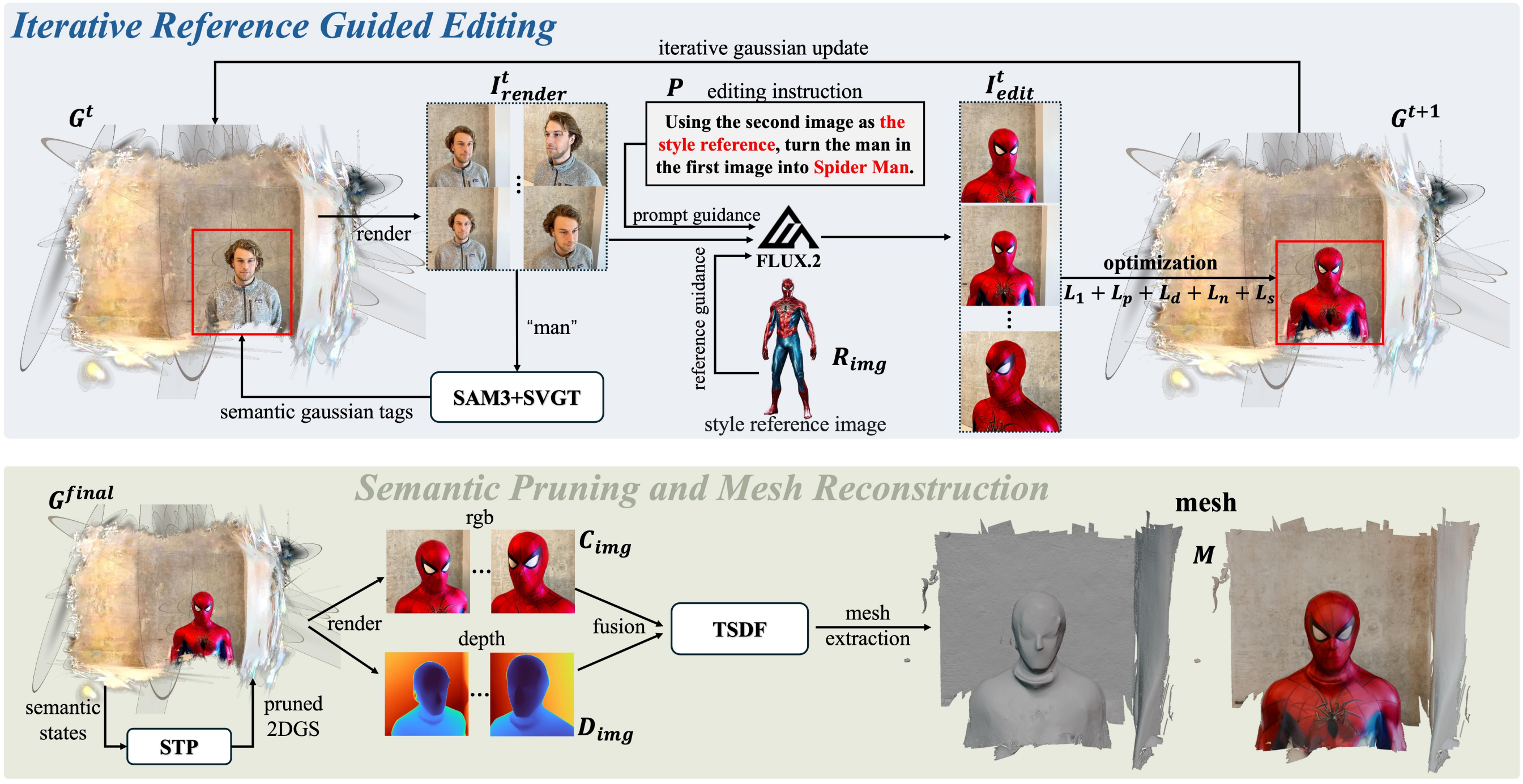}
    \caption{Overview of the Edit2MeshGS pipeline. SVGT first tags the editable Gaussians corresponding to the target region using multi-view masks. Reference-guided FLUX.2 then generates consistent multi-view supervision for geometry-aware 2DGS optimization. After editing, STP removes residual source geometry before refinement and final mesh extraction.}
    \label{fig:arg}
\end{figure*}

\section{Related Work}

\subsection{2D Editing Models}

2D generative priors \cite{ho2020denoising,song2021denoising,rombach2022high} have played an important role in 3D scene editing. Early methods \cite{meng2022sdedit,hertz2023prompt,mokady2023null,couairon2022diffedit,xu2024inversionfree} primarily adopted training-free or lightweight editing strategies to repurpose pretrained diffusion models for text-driven image editing. Training-based methods such as InstructPix2Pix \cite{brooks2023instructpix2pix} and ControlNet \cite{zhang2023adding} further improved instruction-guided editing and spatial structure control, making 2D diffusion priors more convenient for 3D scene optimization. Recent works \cite{flux2024,wu2025qwen,cai2025z,labs2025flux1kontextflowmatching,zhao2026qwen} have further improved image editing through multimodal understanding, precise editing, and efficient diffusion architectures. Meanwhile, video editing models \cite{blattmann2023stable,shi2024motion,ku2024anyvv,jiang2025vace,cao2025instruction} provide temporal consistency priors through temporal modeling or motion constraints, offering new insights for continuous-view editing. However, existing 2D and video editing priors still struggle to simultaneously achieve stable performance in complex instruction following, reference style preservation, and high-fidelity local editing. FLUX.2 \cite{flux-2-2025} builds upon a rectified-flow Transformer generation framework and further integrates vision-language modeling capabilities, thereby demonstrating strong advantages in instruction following, reference image fusion, and high-fidelity editing.

\subsection{3D Scene Editing}

GaussianEditor \cite{chen2024gaussianeditor} first systematically introduced 3D Gaussian Splatting into text-driven 3D scene editing, establishing an important foundation upon which subsequent works further explored multi-view consistent editing. DGE \cite{chen2024dge} and EditSplat \cite{lee2025editsplat} adopt InstructPix2Pix \cite{brooks2023instructpix2pix} as the 2D editing prior and alleviate cross-view inconsistencies caused by independently edited views by integrating multi-view information and 3D geometric cues. GaussCtrl \cite{wu2024gaussctrl} and CoreEditor \cite{zhu2026coreeditor} employ depth-conditioned ControlNet \cite{zhang2023adding} as the editing prior, enhancing geometric preservation in multi-view editing through structure-aware spatial conditioning. VcEdit \cite{wang2024view}, InterGSEdit \cite{wen2025intergsedit}, and TrAME \cite{luo2025trame} embed consistency modeling into diffusion editing trajectories and 3D optimization, improving view consistency and local controllability in progressive 3D editing through cross-view attention modulation. Recently, EditCast3D \cite{qu2025editcast3d}, ViP3DE \cite{chen2026fast}, and I-Mix2Mix \cite{gilo2026instructmix2mix} have further introduced video diffusion and multi-view generative priors, formulating multi-view editing as continuous-view propagation and consistent generation to reduce artifacts caused by per-view editing. Meanwhile, Omni-3DEdit \cite{chen2026omni} and FluSplat \cite{huang2026flusplat} explore sparse-view editing and feed-forward reconstruction, reducing the reliance on long test-time 3D optimization by introducing cross-view consistency learning during model training. However, these methods mainly focus on the consistency of multi-view editing results, while paying insufficient attention to explicit surface structure constraints and mesh extractability of the edited 3D scene, making it difficult to stably reconstruct high-quality meshes.

\section{Method}

\subsection{Overall Pipeline}

Given an initial 2DGS scene $\mathcal{G}^{0}=\{g_i\}_{i=1}^{N}$, an editing instruction $P$, and a style reference image $R_{\mathrm{img}}$, Edit2MeshGS iteratively optimizes $\mathcal{G}^{t}$ into an edited scene that supports both high fidelity rendering and mesh reconstruction, with the overall pipeline shown in Fig.~\ref{fig:arg}. We denote the rendered views and FLUX.2 edited views at iteration $t$ as $\mathcal{I}_{\mathrm{render}}^{t}$ and $\mathcal{I}_{\mathrm{edit}}^{t}$, respectively.

Following 2DGS, each primitive is represented as an oriented planar Gaussian:
\begin{equation}
\mathbf{X}_i(u,v)=\mathbf{p}_i+s_i^u\mathbf{t}_i^u u+s_i^v\mathbf{t}_i^v v,
\end{equation}
where $\mathbf{p}_i$ is the Gaussian center, $\mathbf{t}_i^u,\mathbf{t}_i^v$ are tangent directions, and $s_i^u,s_i^v$ are scales. For pixel $\mathbf{x}$, let $\mathcal{N}(\mathbf{x})=\{i_1,\ldots,i_K\}$ be the projected Gaussians covering this pixel, sorted from front to back. The rendered color is computed by alpha compositing:

\begin{equation}
\hat{\mathbf{I}}(\mathbf{x})
=
\sum_{k=1}^{K}
T_k(\mathbf{x})
\alpha_{i_k}(\mathbf{x})
\mathbf{c}_{i_k},
\quad
T_k(\mathbf{x})
=
\prod_{\ell=1}^{k-1}
\left(1-\alpha_{i_\ell}(\mathbf{x})\right),
\end{equation}
where $\alpha_{i_k}(\mathbf{x})$ and $\mathbf{c}_{i_k}$ are the opacity and view dependent color of Gaussian $g_{i_k}$ at pixel $\mathbf{x}$, and $T_k(\mathbf{x})$ is the accumulated transmittance before the $k$-th Gaussian.

At initialization, we render $\mathcal{I}_{\mathrm{render}}^{t}$ from $\mathcal{G}^{t}$ and use SAM3 \cite{carion2025sam} with the source object prompt to obtain multi-view object masks. SVGT converts these masks into an editable Gaussian subset $\mathcal{G}_{\mathrm{edit}}^{t}\subset\mathcal{G}^{t}$. FLUX.2 then takes $\mathcal{I}_{\mathrm{render}}^{t}$, $R_{\mathrm{img}}$, and $P$ as inputs to generate $\mathcal{I}_{\mathrm{edit}}^{t}$, providing reference guided multi-view editing supervision.

Only $\mathcal{G}_{\mathrm{edit}}^{t}$ is optimized, while the remaining Gaussians are fixed. The optimization objective is:
\begin{equation}
\mathcal{L}_{\mathrm{opt}}
=
\lambda_1\mathcal{L}_{1}
+\lambda_{p}\mathcal{L}_{p}
+\lambda_{d}\mathcal{L}_{d}
+\lambda_{n}\mathcal{L}_{n}
+\lambda_{s}\mathcal{L}_{s}.
\end{equation}
Here, $\mathcal{L}_{1}$ and $\mathcal{L}_{p}$ denote the L1 reconstruction loss and LPIPS perceptual loss for edited view supervision, $\mathcal{L}_{d}$ and $\mathcal{L}_{n}$ are the depth distortion and normal consistency regularizers inherited from 2DGS, and $\mathcal{L}_{s}$ is our surface regularization loss for preserving surface aligned geometry.

For each view $v$ and pixel $\mathbf{x}$, let $D_v^{t}(\mathbf{x})$ be the rendered surface depth and $\mathcal{N}_{v}^{\mathrm{edit}}(\mathbf{x})$ be the editable Gaussians contributing to $\mathbf{x}$. With $\Omega=\{(v,\mathbf{x},i)\mid g_i\in\mathcal{N}_{v}^{\mathrm{edit}}(\mathbf{x})\}$, we define:
\begin{equation}
\mathcal{L}_{s}
=
\frac{1}{|\Omega|}
\sum_{(v,\mathbf{x},i)\in\Omega}
w_{i,v}(\mathbf{x})
\rho\left(
z_{i,v}(\mathbf{x})-D_v^{t}(\mathbf{x})
\right).
\end{equation}

where $z_{i,v}(\mathbf{x})$ is the depth of Gaussian $g_i$ along the ray of pixel $\mathbf{x}$ in view $v$, $D_v^{t}(\mathbf{x})$ is the rendered surface depth, $w_{i,v}(\mathbf{x})$ is the rendering contribution weight, and $\rho(\cdot)$ is a robust penalty. The rendered depth can also be combined with external depth priors when available.

After optimization, STP prunes residual source Gaussians from $\mathcal{G}^{\mathrm{final}}$ and produces $\widetilde{\mathcal{G}}$. Following 2DGS, we render depth maps from $\widetilde{\mathcal{G}}$ over the training views and fuse them with TSDF to extract the mesh $\mathcal{M}$.

\begin{figure}[t]
    \centering
    \includegraphics[width=0.98\columnwidth]{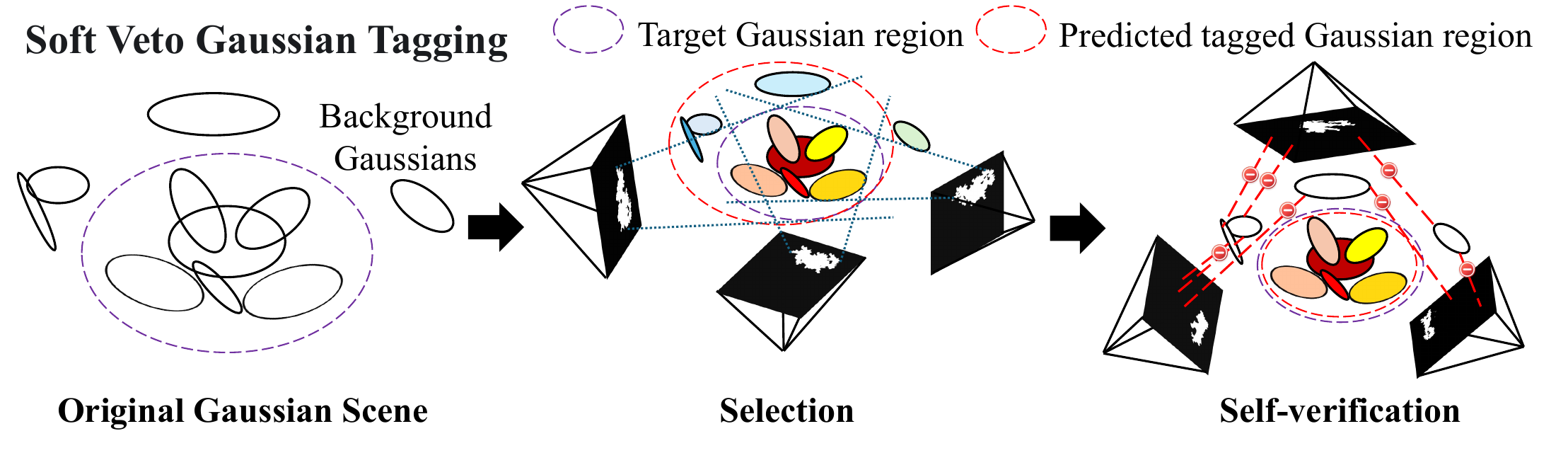}
    \caption{Overview of Soft Veto Gaussian Tagging. Multi-view semantic evidence selects editable Gaussians, and per-view evidence verifies the initial selection.}
    \label{fig:svgt}
\end{figure}

\subsection{Semantic Transition Pruning}

\begin{figure}[t]
    \centering
    \includegraphics[width=0.98\columnwidth]{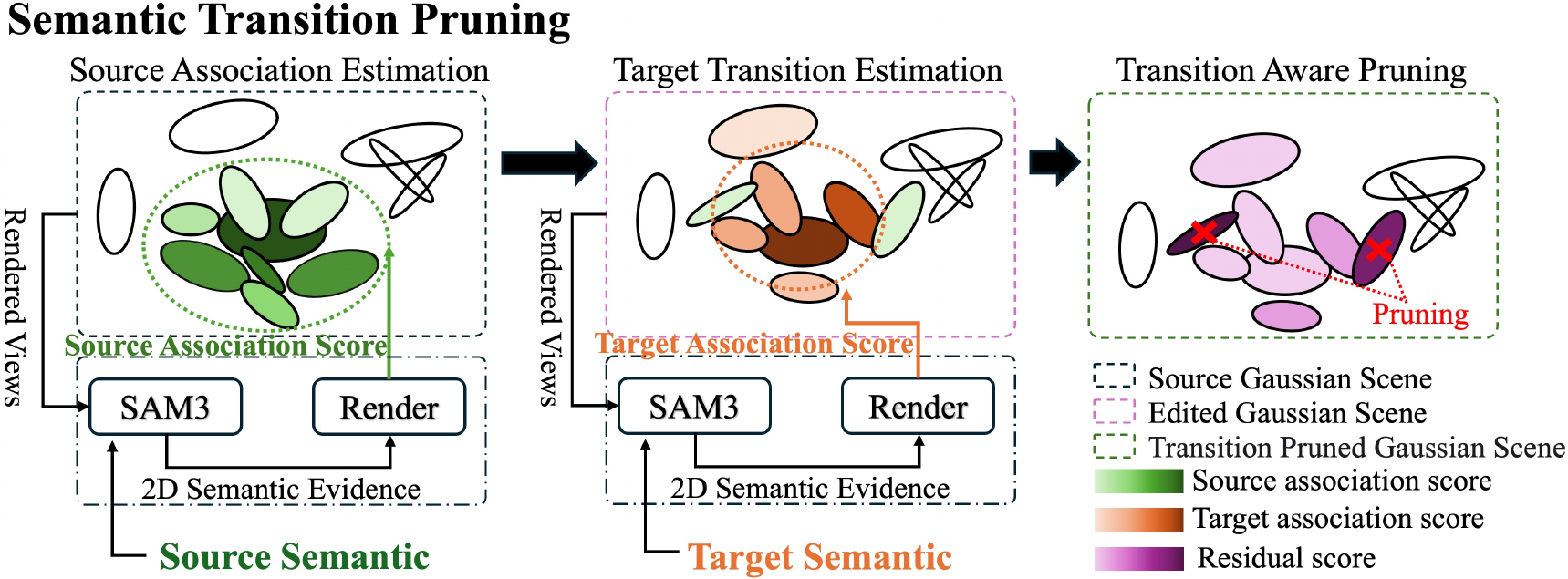}
    \caption{Overview of Semantic Transition Pruning. Source association estimation and target transition estimation are followed by transition-aware Gaussian pruning.}
    \label{fig:stp}
\end{figure}

Large deformation editing may produce visually plausible renderings while leaving source Gaussians hidden by optimized color or opacity. Since these Gaussians still retain geometric primitives, they can be fused into redundant structures during mesh reconstruction. To remove such residual geometry, we introduce Semantic Transition Pruning (STP), which estimates source to target semantic transitions at the Gaussian level and prunes semantically stagnant Gaussians before mesh extraction, as shown in Fig.~\ref{fig:stp}. For a semantic prompt $q$ at stage $t$, SAM3 provides multi-view masks $\{M_v^{q,t}\}_{v\in\mathcal{V}}$. STP defines a continuous semantic association score for each Gaussian $g_i$:
\begin{equation}
p_i^q(t)
=
\frac{
\sum_{v\in\mathcal{V}_i(t)}
w_{i,v}(t)
M_v^{q,t}\left(\pi_v(g_i)\right)
}{
\sum_{v\in\mathcal{V}_i(t)}
w_{i,v}(t)
+\epsilon
},
\end{equation}
where $\mathcal{V}_i(t)$ is the set of views in which $g_i$ is visible, $\pi_v(g_i)$ is its projected position, and $w_{i,v}(t)=\alpha_{i,v}(t)T_{i,v}(t)$ is the alpha compositing contribution weight at the projected position.

We compute the source association before editing and the target association after editing as $s_i=p_i^{q_s}(0)$ and $u_i=p_i^{q_t}(T)$, where $q_s$ and $q_t$ denote the source and target prompts. To account for the potential geometric impact of residual Gaussians, STP uses their final scale products as geometric weights:
\begin{equation}
a_i=s_i^{u}(T)\,s_i^{v}(T),
\end{equation}
where $s_i^{u}(T)$ and $s_i^{v}(T)$ are the scale parameters of Gaussian $g_i$ along its two local tangent directions at the end of editing, respectively. Their product characterizes the Gaussian's surface extent.
The transition residual score is then defined as:
\begin{equation}
r_i=s_i(1-u_i)a_i.
\end{equation}
A high $r_i$ indicates that $g_i$ is strongly associated with the source object, weakly associated with the target object, and still geometrically active in the final scene.

STP ranks the editable Gaussians by their residual scores in descending order and prunes the top $\tau_{\mathrm p}$ fraction:
\begin{equation}
\mathcal{G}_{\mathrm{res}}
=\operatorname{Top}_{\tau_{\mathrm p}}\!\left(\mathcal{G}_{\mathrm{edit}}^{T};r\right),
\quad
\widetilde{\mathcal{G}}
=\mathcal{G}^{\mathrm{final}}\setminus\mathcal{G}_{\mathrm{res}},
\end{equation}
where $\tau_{\mathrm p}$ is the pruning ratio, $\operatorname{Top}_{\tau_{\mathrm p}}(\cdot;r)$ denotes this selection, and $\widetilde{\mathcal{G}}$ is the pruned 2DGS representation. We then briefly refine the remaining Gaussians to reduce local artifacts caused by pruning. In this way, STP removes residual source Gaussians that are hidden in rendered views but would otherwise form redundant geometry during mesh extraction.

\begin{table*}[!t]
    \centering
    \caption{Quantitative comparison over twelve editing tasks. 3DC denotes 3D consistency; US denotes user study scores (\%) from 60 participants. GaussianEditor$^{*}$, Omni-3DEdit$^{*}$, and Edit2MeshGS (klein-9B) share FLUX.2-klein-9B.}
    \label{tab:quantitative-comparison}
    \small
    \setlength{\tabcolsep}{2pt}
    \begin{tabularx}{\textwidth}{@{}>{\raggedright\arraybackslash}p{0.17\textwidth}c*{8}{C}cc@{}}
        \toprule
        \multirow{2}{*}{Method} & \multirow{2}{*}{Pub.} & \multicolumn{4}{c}{Editing Metrics} & \multicolumn{4}{c}{Mesh Metrics} & \multicolumn{2}{c}{Efficiency} \\
        \cmidrule(lr){3-6}\cmidrule(lr){7-10}\cmidrule(l){11-12}
        & & CDS $\uparrow$ & CS $\uparrow$ & 3DC $\uparrow$ & US $\uparrow$ & SCAR $\downarrow$ & NV $\downarrow$ & BER $\downarrow$ & US $\uparrow$ & Time $\downarrow$ & Mem. $\downarrow$ \\
        \midrule
        GaussianEditor~\cite{chen2024gaussianeditor} & \multirow{2}{*}{CVPR 2024} & \rcell{4}{0.1211} & \rcell{4}{0.2287} & \rcell{9}{50.68} & \rcell{6}{1.67} & \rcell{5}{0.0092} & \rcell{3}{51.63} & \rcell{6}{0.0797} & \rcell{4}{10.00} & \rcell{6}{$\sim$20 min} & \rcell{1}{\textbf{$\sim$16 GB}} \\
        GaussianEditor$^{*}$~\cite{chen2024gaussianeditor} &  & \rcell{3}{0.1295} & \rcell{3}{0.2365} & \rcell{3}{59.28} & \rcell{4}{11.67} & \rcell{3}{0.0078} & \rcell{6}{51.96} & \rcell{3}{0.0712} & \rcell{7}{1.67} & \rcell{7}{$\sim$23 min} & \rcell{6}{$\sim$37 GB} \\
        DGE~\cite{chen2024dge} & ECCV 2024 & \rcell{5}{0.1150} & \rcell{6}{0.2237} & \rcell{4}{57.55} & \rcell{3}{13.33} & \rcell{8}{0.0096} & \rcell{5}{51.93} & \rcell{8}{0.0835} & \rcell{3}{11.67} & \rcell{2}{$\sim$6 min} & \rcell{2}{$\sim$20 GB} \\
        VcEdit~\cite{wang2024view} & ECCV 2024 & \rcell{7}{0.1073} & \rcell{5}{0.2239} & \rcell{5}{57.00} & \rcell{6}{1.67} & \rcell{7}{0.0094} & \rcell{4}{51.88} & \rcell{7}{0.0827} & \rcell{5}{6.67} & \rcell{8}{$\sim$32 min} & \rcell{5}{$\sim$28 GB} \\
        EditSplat~\cite{lee2025editsplat} & CVPR 2025 & \rcell{8}{0.0805} & \rcell{7}{0.2146} & \rcell{8}{52.17} & \rcell{5}{5.00} & \rcell{6}{0.0093} & \rcell{8}{52.06} & \rcell{5}{0.0792} & \rcell{6}{3.33} & \rcell{3}{$\sim$8 min} & \rcell{3}{$\sim$21 GB} \\
        VIP3DE~\cite{chen2026fast} & AAAI 2026 & \rcell{9}{0.0195} & \rcell{9}{0.1965} & \rcell{6}{56.38} & \rcell{6}{1.67} & \rcell{4}{0.0089} & \rcell{7}{52.01} & \rcell{4}{0.0787} & \rcell{5}{6.67} & \rcell{5}{$\sim$15 min} & \rcell{9}{$\sim$59 GB} \\
        Omni-3DEdit$^{*}$~\cite{chen2026omni} & CVPR 2026 & \rcell{6}{0.1086} & \rcell{8}{0.2091} & \rcell{7}{56.11} & \rcell{5}{5.00} & \rcell{9}{0.0120} & \rcell{9}{52.17} & \rcell{9}{0.0997} & \rcell{7}{1.67} & \rcell{1}{\textbf{$\sim$4 min}} & \rcell{8}{$\sim$40 GB} \\
        \midrule
        \textbf{Ours (klein-4B)} & -- & \rcell{2}{0.1768} & \rcell{2}{0.2487} & \rcell{2}{70.37} & \rcell{2}{26.67} & \rcell{2}{0.0037} & \rcell{2}{40.66} & \rcell{2}{0.0239} & \rcell{2}{25.00} & \rcell{2}{$\sim$6 min} & \rcell{4}{$\sim$22 GB} \\
        \textbf{Ours (klein-9B)} & -- & \rcell{1}{\textbf{0.1836}} & \rcell{1}{\textbf{0.2584}} & \rcell{1}{\textbf{72.77}} & \rcell{1}{\textbf{33.33}} & \rcell{1}{\textbf{0.0026}} & \rcell{1}{\textbf{39.21}} & \rcell{1}{\textbf{0.0185}} & \rcell{1}{\textbf{33.33}} & \rcell{4}{$\sim$10 min} & \rcell{7}{$\sim$38 GB} \\
        \bottomrule
    \end{tabularx}

\vspace{0.8em}
    \small
    \captionof{table}{Segmentation accuracy (\%) and efficiency across five tasks. IG: InstanceGaussian. Efficiency is averaged over five scenes. Tag Time denotes baseline query time or SVGT tagging time using pre-segmented results.}
    \label{tab:segmentation-comparison}
    \footnotesize
    \setlength{\tabcolsep}{1.5pt}
    \renewcommand{\arraystretch}{1.15}
\begin{tabularx}{\textwidth}{@{}l*{15}{C}cc@{}}
        \toprule
        \multirow{2}{*}{Method}
        & \multicolumn{3}{c}{Face}
        & \multicolumn{3}{c}{Bear}
        & \multicolumn{3}{c}{Doll}
        & \multicolumn{3}{c}{Garden}
        & \multicolumn{3}{c}{Bonsai}
        & \multicolumn{2}{c}{Efficiency} \\
        \cmidrule(lr){2-4}
        \cmidrule(lr){5-7}
        \cmidrule(lr){8-10}
        \cmidrule(lr){11-13}
        \cmidrule(lr){14-16}
        \cmidrule(l){17-18}
        & P & R & mIoU
        & P & R & mIoU
        & P & R & mIoU
        & P & R & mIoU
        & P & R & mIoU
        & Tag Time$\downarrow$ & Mem.$\downarrow$ \\
        \midrule
        SAGA~\cite{cen2025segment}
        & \rcell{2}{98.28} & 98.81 & \rcell{3}{97.12}
        & \rcell{2}{97.86} & \rcell{3}{99.59} & \rcell{2}{97.47}
        & \rcell{2}{98.76} & \rcell{2}{99.69} & \rcell{2}{98.46}
        & \rcell{2}{97.97} & \rcell{3}{93.85} & \rcell{3}{92.05}
        & \rcell{2}{94.99} & 81.13 & \rcell{2}{77.65}
        & \rcell{2}{121~s} & \rcell{3}{$\sim$1.3~GB} \\
        IG~\cite{li2025instancegaussian}
        & 95.44 & \rcell{2}{99.70} & 95.15
        & 91.51 & 99.19 & 90.81
        & 88.10 & \rcell{1}{\textbf{99.88}} & 88.00
        & 94.99 & 92.09 & 87.84
        & \rcell{3}{84.20} & \rcell{2}{90.89} & \rcell{3}{77.62}
        & 142~s & \rcell{2}{$\sim$1.0~GB} \\
        COB-GS~\cite{zhang2025cob}
        & \rcell{1}{\textbf{98.44}} & \rcell{3}{99.61} & \rcell{1}{\textbf{98.06}}
        & \rcell{3}{97.82} & \rcell{2}{99.61} & \rcell{3}{97.45}
        & \rcell{1}{\textbf{98.84}} & \rcell{3}{99.67} & \rcell{1}{\textbf{98.52}}
        & \rcell{1}{\textbf{98.70}} & \rcell{2}{94.88} & \rcell{1}{\textbf{93.70}}
        & 77.58 & \rcell{3}{82.40} & 66.68
        & \rcell{3}{125~s} & $\sim$3.7~GB \\
        \midrule
        \textbf{Ours}
        & \rcell{3}{98.01} & \rcell{1}{\textbf{99.82}} & \rcell{2}{97.84}
        & \rcell{1}{\textbf{98.40}} & \rcell{1}{\textbf{99.86}} & \rcell{1}{\textbf{98.27}}
        & \rcell{3}{97.08} & 95.39 & \rcell{3}{92.29}
        & \rcell{3}{96.53} & \rcell{1}{\textbf{96.03}} & \rcell{2}{92.90}
        & \rcell{1}{\textbf{98.22}} & \rcell{1}{\textbf{95.76}} & \rcell{1}{\textbf{94.42}}
        & \rcell{1}{\textbf{7~s}} & \rcell{1}{\textbf{$\sim$0.9~GB}} \\
        \bottomrule
    \end{tabularx}

    \vspace{16pt}
    \small
    \begin{minipage}[t]{0.57\textwidth}
        \centering
        \captionsetup{skip=4pt}
        \captionof{table}{Quantitative ablation of Edit2MeshGS over twelve editing tasks. 3DC denotes 3D consistency.}
        \label{tab:mesh-ablation}
        \small
        \setlength{\tabcolsep}{1.5pt}
        \renewcommand{\arraystretch}{0.835}
        \begin{tabularx}{\linewidth}{@{}l*{6}{C}@{}}
            \toprule
            Setting & CDS $\uparrow$ & CS $\uparrow$ & 3DC $\uparrow$ & SCAR $\downarrow$ & NV $\downarrow$ & BER $\downarrow$ \\
            \midrule
            3DGS + IP2P & \rcell{7}{0.0897} & \rcell{7}{0.2131} & \rcell{7}{51.22} & \rcell{8}{0.0090} & \rcell{8}{51.98} & \rcell{8}{0.0797} \\
            3DGS + FLUX.2 & \rcell{6}{0.1322} & \rcell{6}{0.2411} & \rcell{6}{58.39} & \rcell{7}{0.0088} & \rcell{7}{51.93} & \rcell{7}{0.0801} \\
            3DGS + FLUX.2 ref. & 0.1670 & 0.2473 & 64.10 & 0.0051 & 46.25 & 0.0418\\
            \midrule
            2DGS + IP2P & \rcell{8}{0.0810} & \rcell{8}{0.2105} & \rcell{8}{51.13} & \rcell{4}{0.0036} & \rcell{4}{43.56} & \rcell{4}{0.0251} \\
            2DGS + FLUX.2 & \rcell{5}{0.1449} & \rcell{5}{0.2395} & \rcell{5}{58.85} & \rcell{6}{0.0063} & \rcell{6}{47.32} & \rcell{6}{0.0411} \\
            2DGS + FLUX.2 ref. & \rcell{4}{0.1823} & \rcell{4}{0.2553} & \rcell{4}{67.71} & \rcell{5}{0.0042} & \rcell{5}{44.69} & \rcell{5}{0.0300} \\
            + SVGT & \rcell{2}{0.1904} & \rcell{3}{0.2582} & \rcell{3}{69.97} & \rcell{3}{0.0035} & \rcell{3}{43.14} & \rcell{2}{0.0235} \\
            + $\mathcal{L}_{s}$ & \rcell{1}{\textbf{0.1914}} & \rcell{1}{\textbf{0.2592}} & \rcell{2}{71.31} & \rcell{2}{0.0033} & \rcell{2}{40.89} & \rcell{3}{0.0236} \\
            Full (+ STP) & \rcell{3}{0.1836} & \rcell{2}{0.2584} & \rcell{1}{\textbf{72.77}} & \rcell{1}{\textbf{0.0026}} & \rcell{1}{\textbf{39.21}} & \rcell{1}{\textbf{0.0185}} \\
            \bottomrule
        \end{tabularx}
    \end{minipage}
    \hfill
    \begin{minipage}[t]{0.41\textwidth}
        \centering
        \captionsetup{skip=4pt}
        \captionof{table}{Quantitative ablation of SVGT over five editing cases. GaussianEditor tracing uses 40 mask views.}
        \label{tab:mask-tagging}
        \small
        \setlength{\tabcolsep}{0.75pt}
        \renewcommand{\arraystretch}{1.11}
        \begin{tabularx}{\linewidth}{@{}l*{6}{C}@{}}
            \toprule
            Setting & $|\mathcal{V}^{+}|$ & $|\mathcal{V}^{-}|$ & $\lambda_{\mathrm{veto}}$ & mIoU$\uparrow$ & P$\uparrow$ & R$\uparrow$ \\
            \midrule
            GE tracing & -- & -- & -- & \rcell{6}{93.05} & \rcell{3}{96.70} & \rcell{7}{94.45} \\
            SVGT-1 & 20 & 5 & 2 & \rcell{3}{94.45} & \rcell{2}{97.16} & \rcell{4}{95.82} \\
            SVGT-2 & 20 & 20 & 5 & \rcell{1}{\textbf{95.14}} & \rcell{1}{\textbf{97.65}} & \rcell{2}{97.37} \\
            SVGT-3 & 10 & 5 & 3.5 & \rcell{2}{94.87} & \rcell{7}{95.26} & \rcell{1}{\textbf{97.78}} \\
            SVGT-4 & 20 & 10 & 3.5 & \rcell{5}{93.40} & \rcell{4}{96.27} & \rcell{5}{94.64} \\
            SVGT-5 & 30 & 20 & 3.5 & \rcell{7}{92.78} & \rcell{5}{95.80} & \rcell{6}{94.48} \\
            SVGT-6 & 30 & 10 & 2 & \rcell{4}{94.37} & \rcell{6}{95.39} & \rcell{3}{97.20} \\
            \bottomrule
        \end{tabularx}
    \end{minipage}
\end{table*}

\subsection{Soft Veto Gaussian Tagging}

Reliable local Gaussian masks are essential for local scene editing. DGE and EditSplat \cite{chen2024dge,lee2025editsplat} use GaussianEditor's semantic tracing \cite{chen2024gaussianeditor}, which can mislabel background Gaussians when object and background projections overlap. Inspired by COB-GS, SAGA, and InstanceGaussian \cite{zhang2025cob,cen2025segment,li2025instancegaussian}, we observe that 2D semantic evidence can also verify the 3D Gaussian masks. We propose self-verifying Soft Veto Gaussian Tagging (SVGT). SVGT first aggregates semantic evidence from one set of views to select editable Gaussians, then checks them against per-view semantic evidence from another set. It converts disagreements into veto penalties to suppress background false positives. Fig.~\ref{fig:svgt} illustrates the overall process.

Given the rendered views $\mathcal{V}$ of the current 2DGS scene, we use SAM3 to generate object masks $\{M_v\}_{v\in\mathcal{V}}$ from a prompt describing the object to be edited. For each Gaussian $g_i$ visible in view $v$, we define its per-view semantic evidence as
\begin{equation}
h_{i,v}=\mathbf{1}[M_v(\pi_v(g_i))=1],
\end{equation}
where $\pi_v(g_i)$ is the projected position of $g_i$. A value of $h_{i,v}=1$ indicates that the view supports including $g_i$ in the editable set.

\begin{figure*}[!t]
    \centering
    \includegraphics[width=0.98\textwidth]{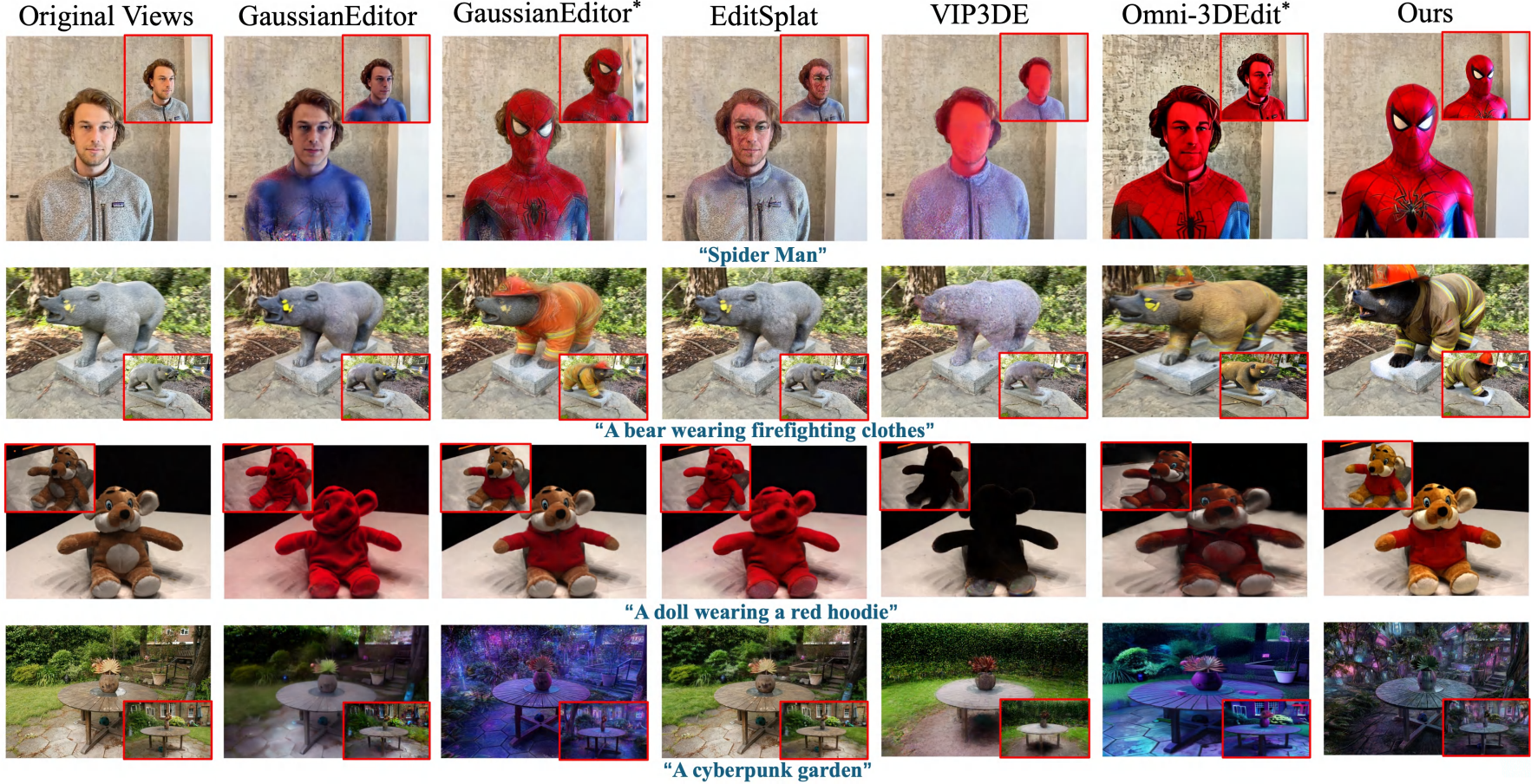}
    \caption{Qualitative comparison of rendered views produced by different 3D scene editing methods. GaussianEditor$^{*}$, Omni-3DEdit$^{*}$, and Ours (klein-9B) all use FLUX.2-klein-9B as the editing model for a fair comparison.}
    \label{fig:xg-comp}
\end{figure*}

\begin{figure*}[!t]
    \centering
    \includegraphics[width=0.98\textwidth]{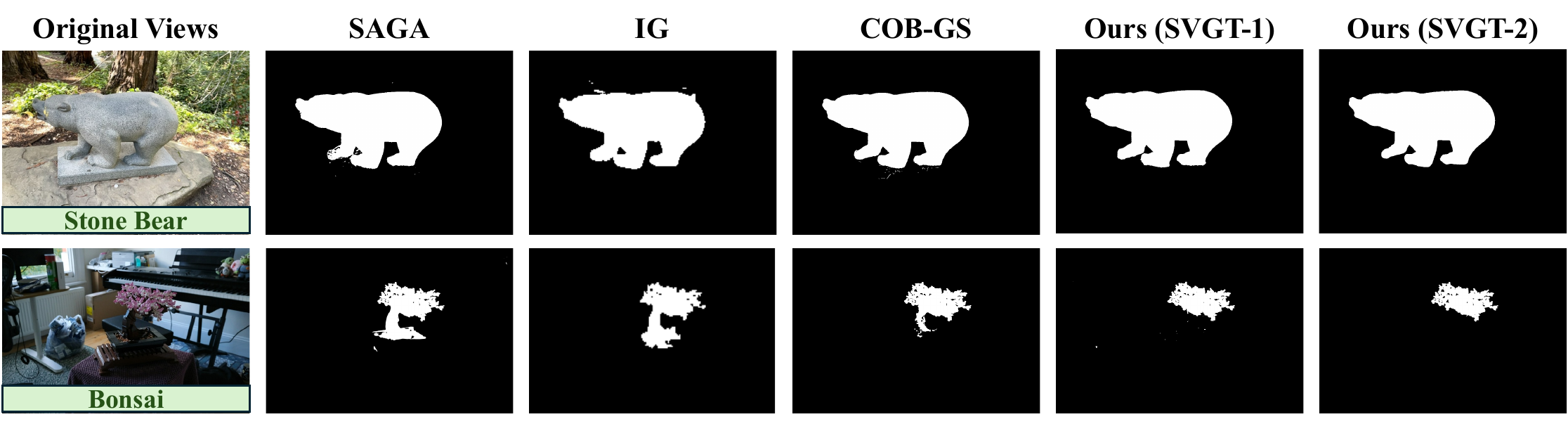}
    \caption{Qualitative comparison of Gaussian segmentation on the Stone Bear and Bonsai scenes. IG denotes InstanceGaussian.}
    \label{fig:seg-comp}
\end{figure*}

Let $\mathcal{V}^{+}$ and $\mathcal{V}^{-}$ denote the view sets for initial selection and verification, respectively. Their sizes are $|\mathcal{V}^{+}|$ and $|\mathcal{V}^{-}|$. The corresponding subsets in which $g_i$ is visible are denoted by $\mathcal{V}_i^{+}$ and $\mathcal{V}_i^{-}$. SVGT aggregates semantic evidence over $\mathcal{V}_i^{+}$ to compute the initial support score
\begin{equation}
s_i^{+}=\frac{1}{|\mathcal{V}_i^{+}|}
\sum_{v\in\mathcal{V}_i^{+}}h_{i,v}.
\end{equation}
We rank the Gaussians by $s_i^{+}$ and retain the top $\tau$ fraction to form the initial editable set $\mathcal{C}$.

SVGT then checks each $g_i\in\mathcal{C}$ against per-view semantic evidence from $\mathcal{V}_i^{-}$. A projection outside the object mask disagrees with the initial selection, as indicated by $1-h_{i,v}$. SVGT penalizes the fraction of views that disagree to obtain the refined score
\begin{equation}
s_i^{\mathrm{tag}}=s_i^{+}
-\lambda_{\mathrm{veto}}
\frac{1}{|\mathcal{V}_i^{-}|}
\sum_{v\in\mathcal{V}_i^{-}}(1-h_{i,v}),
\qquad g_i\in\mathcal{C},
\end{equation}
where $\lambda_{\mathrm{veto}}$ is the veto weight. Candidates whose refined scores reach the threshold $\theta_{\mathrm{tag}}$ form the final editable set:
\begin{equation}
\mathcal{G}_{\mathrm{edit}}
=\{g_i\in\mathcal{C}\mid s_i^{\mathrm{tag}}\ge\theta_{\mathrm{tag}}\}.
\end{equation}

\begin{figure*}[!t]
    \centering
    \includegraphics[width=1\textwidth]{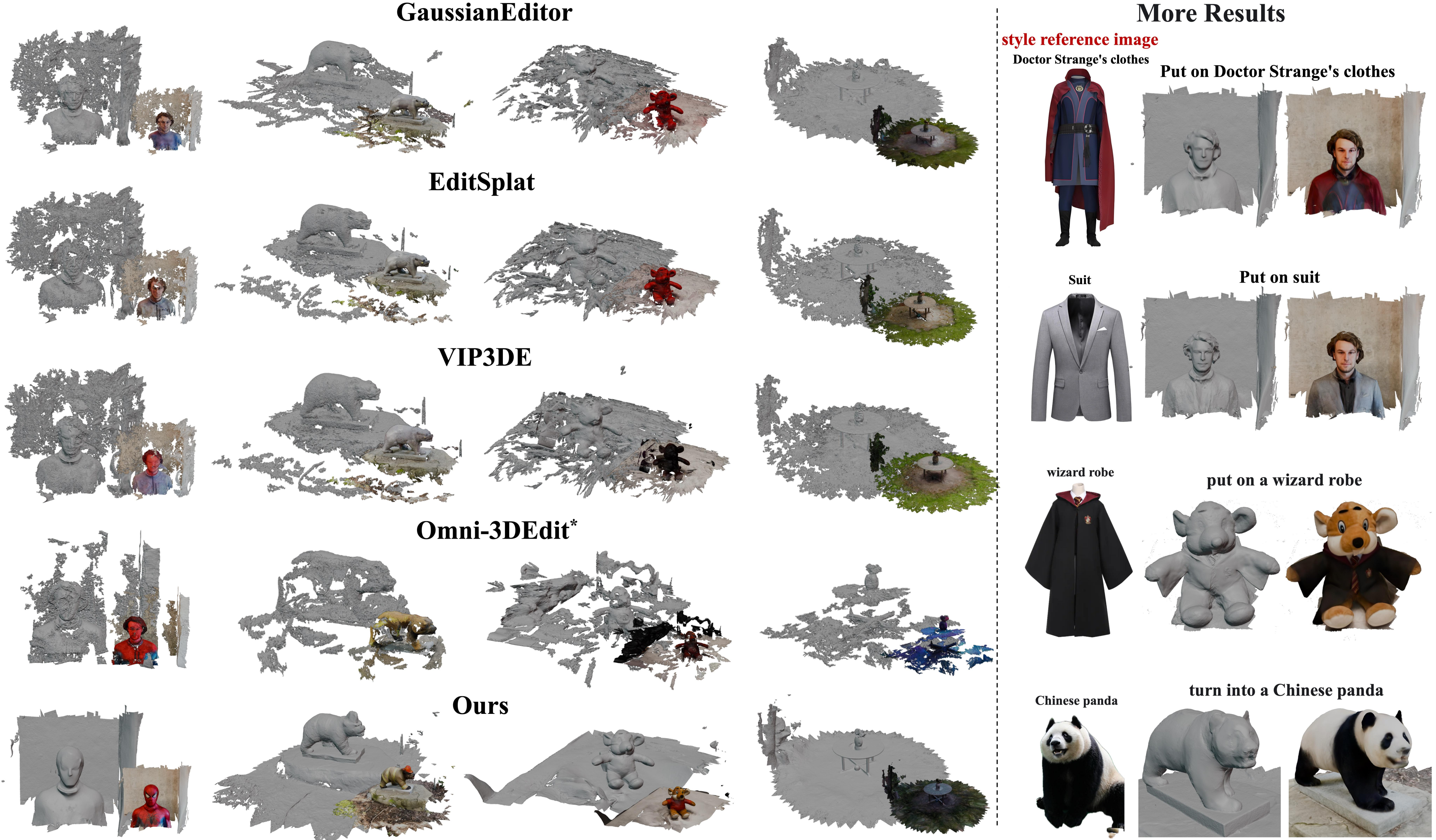}
    \caption{Qualitative comparison of reconstructed meshes. The left side compares different 3D scene editing methods, and the right side shows additional mesh reconstruction results of Edit2MeshGS.}
    \label{fig:mesh-comp}
\end{figure*}

\section{Experiments}

\subsection{Experimental Details}

We quantitatively compare Edit2MeshGS with GaussianEditor \cite{chen2024gaussianeditor}, GaussianEditor$^{*}$ \cite{chen2024gaussianeditor}, DGE \cite{chen2024dge}, VcEdit \cite{wang2024view}, EditSplat \cite{lee2025editsplat}, VIP3DE \cite{chen2026fast}, and Omni-3DEdit$^{*}$ \cite{chen2026omni}, and select representative methods for qualitative comparison. We compare SVGT with three Gaussian segmentation methods: SAGA \cite{cen2025segment}, InstanceGaussian \cite{li2025instancegaussian}, and COB-GS \cite{zhang2025cob}. For a consistent comparison, we use SAM3 to generate the 2D segmentation masks for all methods. GaussianEditor, DGE, and EditSplat use InstructPix2Pix \cite{brooks2023instructpix2pix} as the editing prior; VcEdit uses InfEdit \cite{xu2024inversionfree}; GaussianEditor$^{*}$, Omni-3DEdit$^{*}$ and our method use FLUX.2-klein-9B \cite{flux-2-2025}; and VIP3DE uses SVD-XT \cite{blattmann2023stable}. For mesh reconstruction, we follow the 2DGS mesh extraction protocol by rendering depth maps from training views and fusing them with TSDF using Open3D. We extract meshes from all other 3DGS-based methods in the same manner for a fair comparison. The voxel size is set to $0.004$, and the truncation threshold is set to $0.02$. All experiments are conducted on a single NVIDIA L20 GPU.

Experiments use scenes from Mip-NeRF 360 \cite{barron2022mip}, Instruct-NeRF2NeRF \cite{haque2023instruct}, and DreamEditor \cite{zhuang2023dreameditor}. For Gaussian segmentation, we report P, R, and mIoU as the average precision, recall, and intersection over union across all views, respectively. The pseudo-labels used for metric computation are generated by SAM3 with additional point prompts. Following prior 3D editing works \cite{chen2024dge,lee2025editsplat}, we evaluate edited renderings using CLIP Directional Similarity (CDS) and CLIP Similarity (CS), which measure editing direction consistency and target semantic alignment. Following GroupEditing \cite{ma2026group}, we additionally use 3D Consistency (3DC) to evaluate multi-view consistency. We evaluate reconstructed meshes using Small Component Area Ratio (SCAR), Normal Variation (NV), and Boundary Edge Ratio (BER), which measure floating fragments and residual geometry, local surface roughness, and holes or open boundaries. Lower mesh metrics indicate cleaner, smoother, and more continuous meshes. We additionally conduct a user study (US) with 60 participants to evaluate the rendering fidelity and mesh quality of the edited results. All metrics are averaged over twelve editing tasks. See the supplementary material for details.

\subsection{Quantitative Comparison}
Table~\ref{tab:quantitative-comparison} reports the quantitative comparison with existing 3D scene editing methods. For edited rendering quality, Edit2MeshGS achieves the highest CDS, CS, and 3DC, together with the highest rendering US score, demonstrating strong semantic alignment, multi-view consistency, and rendering fidelity.

\begin{figure}[!t]
    \centering
    \includegraphics[width=1\columnwidth]{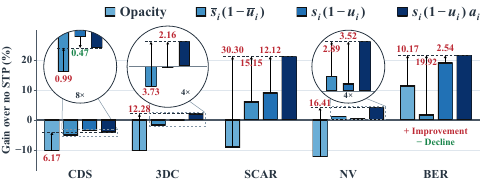}
    \caption{Comparison of Gaussian pruning strategies. $\bar{s}_i$ and $\bar{u}_i$ denote source and target mask associations obtained by unweighted averaging over their respective visible views. Bars show relative gains (\%) over no STP; higher is better.}
    \label{fig:xr-stp}
\end{figure}

For mesh reconstruction quality, Edit2MeshGS achieves the lowest SCAR, NV, and BER, together with the highest mesh US score. These results indicate fewer floating fragments, smoother local surfaces, and fewer open boundaries, consistent with participants' preferences. The ablations in Table~\ref{tab:mesh-ablation} and Figs.~\ref{fig:xr-stp} and~\ref{fig:xr-mesh} further demonstrate the contributions of surface regularization and STP to mesh reconstruction quality. Edit2MeshGS requires approximately 10 minutes and 38 GB of memory.

As shown in Table~\ref{tab:segmentation-comparison} and Fig.~\ref{fig:seg-comp}, SVGT achieves highly efficient Gaussian tagging through a lightweight improvement and remains competitive with existing Gaussian segmentation methods.
\subsection{Qualitative Comparisons}
Fig.~\ref{fig:xg-comp} shows the qualitative comparison of rendered views. GaussianEditor and EditSplat often fail to follow complex instructions due to the limited editing ability of InstructPix2Pix. Although GaussianEditor$^{*}$ benefits from the stronger editing capability of FLUX.2, independently editing each view introduces cross-view inconsistencies and visible artifacts. VIP3DE tends to preserve the original appearance because of its video prior, while Omni-3DEdit$^{*}$ remains less effective for character replacement and local clothing editing. In contrast, Edit2MeshGS uses FLUX.2 with a style reference image and precise text instruction, producing faithful and multi-view consistent edits across different scenes.

Fig.~\ref{fig:mesh-comp} shows the qualitative comparison of reconstructed meshes. Existing 3DGS based methods mainly focus on multi-view consistent edited renderings and lack sufficient geometry constraints for mesh reconstruction, leading to noisy surfaces, holes, or redundant structures. In contrast, Edit2MeshGS combines surface regularization with STP to improve surface smoothness and reduce residual source geometry, producing cleaner reconstructed meshes. Additional results on the right side of Fig.~\ref{fig:mesh-comp} further show that our method can generate both high quality rendered views and usable mesh assets.

\subsection{Ablation Study}
Table~\ref{tab:mesh-ablation} reports the contribution of each component. Replacing 3DGS with 2DGS improves the mesh metrics. Using FLUX.2 improves CDS and CS, but degrades mesh quality compared with 2DGS + IP2P, because its larger appearance changes and residual cross-view inconsistencies lead to stronger and less stable Gaussian updates without sufficient regularization. Reference guidance alleviates this issue and improves both editing and mesh metrics. SVGT further improves all metrics by producing more reliable local Gaussian masks, and $\mathcal{L}_{s}$ mainly improves surface related metrics. STP further improves mesh metrics, although CDS and CS slightly decrease due to minor rendering artifacts introduced by direct Gaussian pruning. Fig.~\ref{fig:xr-stp} compares low-opacity pruning (Opacity), unweighted mask-based scoring $\bar{s}_i(1-\bar{u}_i)$, rendering-contribution-weighted scoring $s_i(1-u_i)$, and full STP $s_i(1-u_i)a_i$. Full STP achieves the highest 3DC and the lowest SCAR, NV, and BER among these strategies, while maintaining a CDS close to that of weighted association alone.

\begin{figure}[!t]
    \centering
    \includegraphics[width=1\columnwidth]{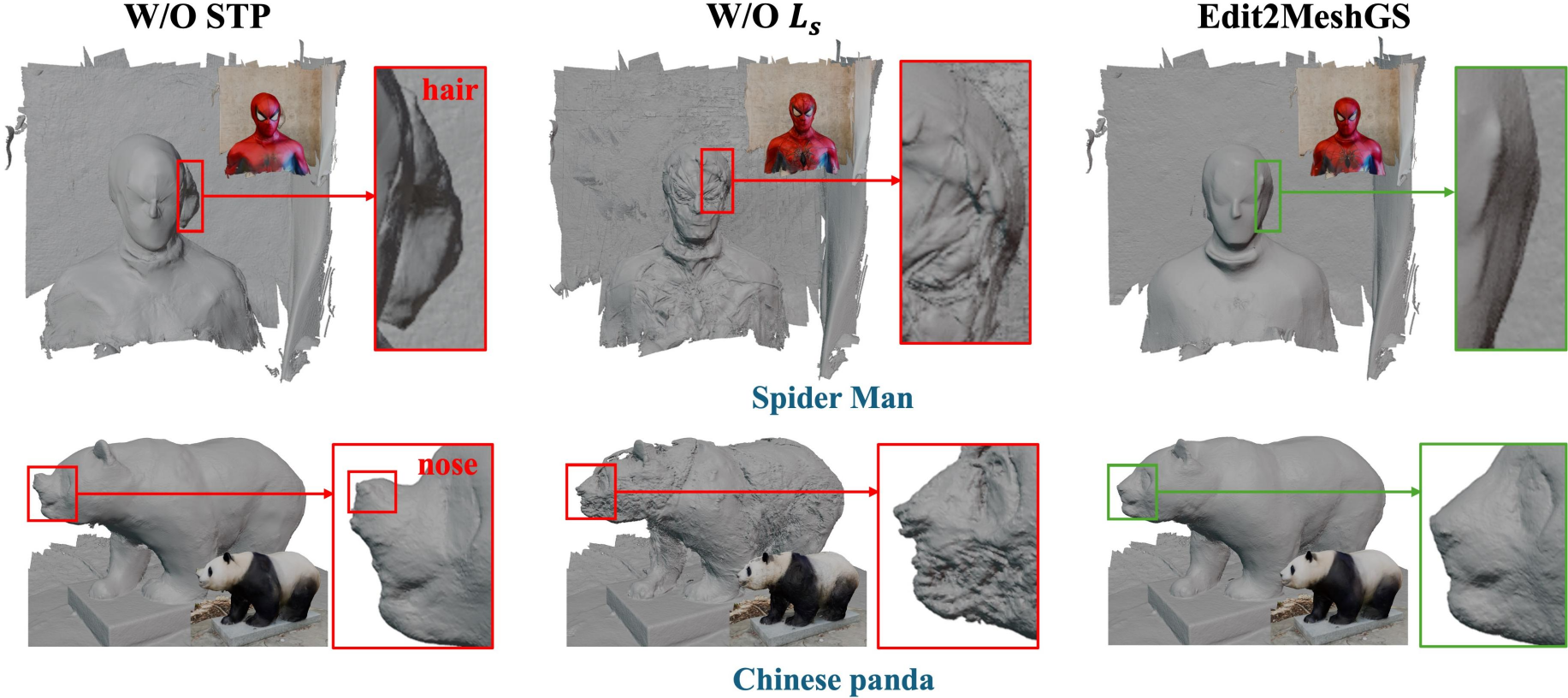}
    \caption{Qualitative ablation of $\mathcal{L}_s$ and STP. Without STP, source geometry remains; without $\mathcal{L}_s$, surfaces become rough. The full method produces cleaner, smoother meshes.}
    \label{fig:xr-mesh}
\end{figure}

\begin{figure}[t]
    \centering
    \includegraphics[width=1\columnwidth]{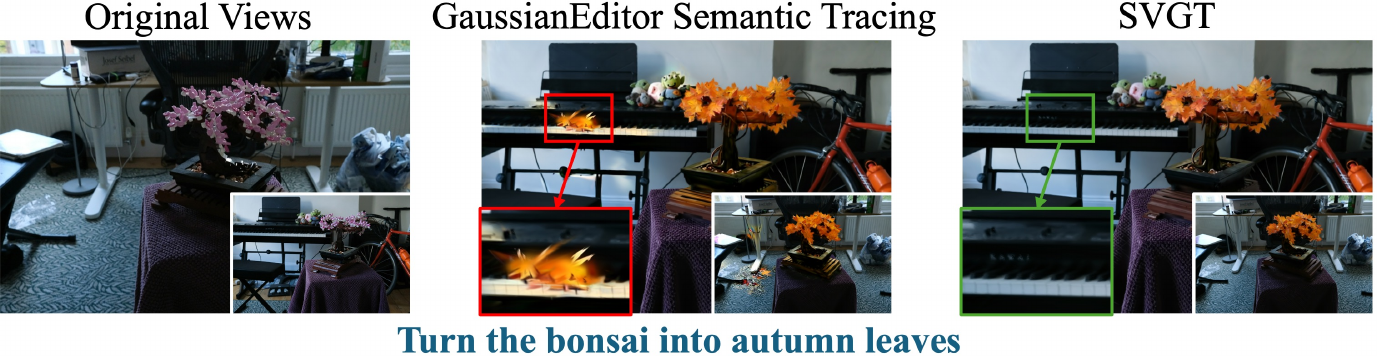}
    \caption{Qualitative ablation of SVGT. GaussianEditor semantic tracing introduces background leaf artifacts (red), while SVGT preserves the highlighted piano region (green).}
    \label{fig:xr-svgt}
\end{figure}

Fig.~\ref{fig:xr-svgt} and Table~\ref{tab:mask-tagging} evaluate SVGT. With 20 views for initial selection and 20 for verification, SVGT-2 reduces background false positives and achieves higher mIoU, precision, and recall than GaussianEditor tracing with 40 mask views. Fig.~\ref{fig:xr-mesh} evaluates $\mathcal{L}_{s}$ and STP for mesh reconstruction. $\mathcal{L}_{s}$ produces smoother surfaces, while STP removes visually hidden residual geometry, such as the original hair of the man and the nose of the stone bear, improving mesh usability with results consistent with Table~\ref{tab:mesh-ablation}. See the supplementary material for details.

\section{Limitation}

Edit2MeshGS requires users to provide a style reference image and appropriately designed prompts to achieve better editing results. In complex editing scenarios, there is still room to improve the stability of semantic transition estimation. Simplifying editing guidance and enhancing the robustness of semantic transition estimation are directions worth further exploration.

\section{Conclusion}

In this paper, we propose Edit2MeshGS, a 2DGS-based framework that unifies high-fidelity 3D scene editing and reliable mesh reconstruction. Shared conditioning guides a diffusion-based editing model to align editing directions across views and provide high-fidelity editing supervision, while geometric constraints regularize local surface geometry to mitigate geometric degradation. Semantic Transition Pruning models source-to-target semantic transitions to remove residual source geometry, while Soft Veto Gaussian Tagging enables local editing and preserves the geometry of unedited regions. Experiments demonstrate improvements of 41.8\% in CLIP score and 74.0\% in mesh reconstruction quality over the second-best method.

\clearpage
\section*{LLM Usage Statement}
OpenAI Codex was used to assist with manuscript editing, LaTeX formatting, and the preparation of analysis and visualization scripts. The authors retain responsibility for the scientific claims, reported results, code verification, and final manuscript content.

{
    \small
    \bibliographystyle{ieeenat_fullname}
    \bibliography{main}

\begin{thebibliography}{51}
\providecommand{\natexlab}[1]{#1}
\providecommand{\url}[1]{\texttt{#1}}
\expandafter\ifx\csname urlstyle\endcsname\relax
  \providecommand{\doi}[1]{doi: #1}\else
  \providecommand{\doi}{doi: \begingroup \urlstyle{rm}\Url}\fi

\bibitem[Barda et~al.(2025)Barda, Gadelha, Kim, Aigerman, Bermano, and
  Groueix]{barda2025instant3dit}
Amir Barda, Matheus Gadelha, Vladimir~G Kim, Noam Aigerman, Amit~H Bermano, and
  Thibault Groueix.
\newblock Instant3dit: Multiview inpainting for fast editing of 3d objects.
\newblock In \emph{Proceedings of the Computer Vision and Pattern Recognition
  Conference}, pages 16273--16282, 2025.

\bibitem[Barron et~al.(2022)Barron, Mildenhall, Verbin, Srinivasan, and
  Hedman]{barron2022mip}
Jonathan~T Barron, Ben Mildenhall, Dor Verbin, Pratul~P Srinivasan, and Peter
  Hedman.
\newblock Mip-nerf 360: Unbounded anti-aliased neural radiance fields.
\newblock In \emph{Proceedings of the IEEE/CVF conference on computer vision
  and pattern recognition}, pages 5470--5479, 2022.

\bibitem[Blattmann et~al.(2023)Blattmann, Dockhorn, Kulal, Mendelevitch,
  Kilian, Lorenz, Levi, English, Voleti, Letts, et~al.]{blattmann2023stable}
Andreas Blattmann, Tim Dockhorn, Sumith Kulal, Daniel Mendelevitch, Maciej
  Kilian, Dominik Lorenz, Yam Levi, Zion English, Vikram Voleti, Adam Letts,
  et~al.
\newblock Stable video diffusion: Scaling latent video diffusion models to
  large datasets.
\newblock \emph{arXiv preprint arXiv:2311.15127}, 2023.

\bibitem[Brooks et~al.(2023)Brooks, Holynski, and
  Efros]{brooks2023instructpix2pix}
Tim Brooks, Aleksander Holynski, and Alexei~A Efros.
\newblock Instructpix2pix: Learning to follow image editing instructions.
\newblock In \emph{Proceedings of the IEEE/CVF conference on computer vision
  and pattern recognition}, pages 18392--18402, 2023.

\bibitem[Cai et~al.(2025)Cai, Cao, Du, Gao, Hoi, Hou, Huang, Jiang, Jin, Li,
  et~al.]{cai2025z}
Huanqia Cai, Sihan Cao, Ruoyi Du, Peng Gao, Steven Hoi, Zhaohui Hou, Shijie
  Huang, Dengyang Jiang, Xin Jin, Liangchen Li, et~al.
\newblock Z-image: An efficient image generation foundation model with
  single-stream diffusion transformer.
\newblock \emph{arXiv preprint arXiv:2511.22699}, 2025.

\bibitem[Cao et~al.(2025)Cao, Zhang, Zheng, and Xia]{cao2025instruction}
Mingdeng Cao, Xuaner Zhang, Yinqiang Zheng, and Zhihao Xia.
\newblock Instruction-based image manipulation by watching how things move.
\newblock In \emph{Proceedings of the Computer Vision and Pattern Recognition
  Conference}, pages 2704--2713, 2025.

\bibitem[Carion et~al.(2025)Carion, Gustafson, Hu, Debnath, Hu, Suris, Ryali,
  Alwala, Khedr, Huang, et~al.]{carion2025sam}
Nicolas Carion, Laura Gustafson, Yuan-Ting Hu, Shoubhik Debnath, Ronghang Hu,
  Didac Suris, Chaitanya Ryali, Kalyan~Vasudev Alwala, Haitham Khedr, Andrew
  Huang, et~al.
\newblock Sam 3: Segment anything with concepts.
\newblock \emph{arXiv preprint arXiv:2511.16719}, 2025.

\bibitem[Cen et~al.(2025)Cen, Fang, Yang, Xie, Zhang, Shen, and
  Tian]{cen2025segment}
Jiazhong Cen, Jiemin Fang, Chen Yang, Lingxi Xie, Xiaopeng Zhang, Wei Shen, and
  Qi Tian.
\newblock Segment any 3d gaussians.
\newblock In \emph{Proceedings of the AAAI conference on artificial
  intelligence}, pages 1971--1979, 2025.

\bibitem[Chen et~al.(2026{\natexlab{a}})Chen, Li, Zhang, Wang, and
  Zhang]{chen2026fast}
Liyi Chen, Ruihuang Li, Guowen Zhang, Pengfei Wang, and Lei Zhang.
\newblock Fast multi-view consistent 3d editing with video priors.
\newblock In \emph{Proceedings of the AAAI Conference on Artificial
  Intelligence}, pages 2948--2956, 2026{\natexlab{a}}.

\bibitem[Chen et~al.(2026{\natexlab{b}})Chen, Wang, Zhang, Ma, and
  Zhang]{chen2026omni}
Liyi Chen, Pengfei Wang, Guowen Zhang, Zhiyuan Ma, and Lei Zhang.
\newblock Omni-3dedit: Generalized versatile 3d editing in one-pass.
\newblock In \emph{Proceedings of the IEEE/CVF Conference on Computer Vision
  and Pattern Recognition}, pages 12640--12650, 2026{\natexlab{b}}.

\bibitem[Chen et~al.(2024{\natexlab{a}})Chen, Laina, and Vedaldi]{chen2024dge}
Minghao Chen, Iro Laina, and Andrea Vedaldi.
\newblock {DGE}: Direct gaussian 3d editing by consistent multi-view editing.
\newblock In \emph{Proceedings of the European Conference on Computer Vision},
  pages 74--92. Springer, 2024{\natexlab{a}}.

\bibitem[Chen et~al.(2024{\natexlab{b}})Chen, Chen, Zhang, Wang, Yang, Wang,
  Cai, Yang, Liu, and Lin]{chen2024gaussianeditor}
Yiwen Chen, Zilong Chen, Chi Zhang, Feng Wang, Xiaofeng Yang, Yikai Wang,
  Zhongang Cai, Lei Yang, Huaping Liu, and Guosheng Lin.
\newblock Gaussianeditor: Swift and controllable 3d editing with gaussian
  splatting.
\newblock In \emph{Proceedings of the IEEE/CVF conference on computer vision
  and pattern recognition}, pages 21476--21485, 2024{\natexlab{b}}.

\bibitem[Couairon et~al.(2022)Couairon, Verbeek, Schwenk, and
  Cord]{couairon2022diffedit}
Guillaume Couairon, Jakob Verbeek, Holger Schwenk, and Matthieu Cord.
\newblock Diffedit: Diffusion-based semantic image editing with mask guidance.
\newblock \emph{arXiv preprint arXiv:2210.11427}, 2022.

\bibitem[Gao et~al.(2023)Gao, Aigerman, Groueix, Kim, and
  Hanocka]{gao2023textdeformer}
William Gao, Noam Aigerman, Thibault Groueix, Vova Kim, and Rana Hanocka.
\newblock Textdeformer: Geometry manipulation using text guidance.
\newblock In \emph{ACM SIGGRAPH 2023 conference proceedings}, pages 1--11,
  2023.

\bibitem[Gilo and Litany(2026)]{gilo2026instructmix2mix}
Daniel Gilo and Or Litany.
\newblock Instructmix2mix: Consistent sparse-view editing through multi-view
  model personalization.
\newblock In \emph{Proceedings of the IEEE/CVF Conference on Computer Vision
  and Pattern Recognition}, pages 29000--29011, 2026.

\bibitem[Haque et~al.(2023)Haque, Tancik, Efros, Holynski, and
  Kanazawa]{haque2023instruct}
Ayaan Haque, Matthew Tancik, Alexei~A Efros, Aleksander Holynski, and Angjoo
  Kanazawa.
\newblock Instruct-nerf2nerf: Editing 3d scenes with instructions.
\newblock In \emph{Proceedings of the IEEE/CVF international conference on
  computer vision}, pages 19740--19750, 2023.

\bibitem[He et~al.(2024)He, Huang, Nie, Hui, Liu, Dai, Han, Li, and
  Liu]{he2024customize}
Runze He, Shaofei Huang, Xuecheng Nie, Tianrui Hui, Luoqi Liu, Jiao Dai,
  Jizhong Han, Guanbin Li, and Si Liu.
\newblock Customize your nerf: Adaptive source driven 3d scene editing via
  local-global iterative training.
\newblock In \emph{Proceedings of the IEEE/CVF conference on computer vision
  and pattern recognition}, pages 6966--6975, 2024.

\bibitem[Hertz et~al.(2023)Hertz, Mokady, Tenenbaum, Aberman, Pritch, and
  Cohen-Or]{hertz2023prompt}
Amir Hertz, Ron Mokady, Jay Tenenbaum, Kfir Aberman, Yael Pritch, and Daniel
  Cohen-Or.
\newblock Prompt-to-prompt image editing with cross-attention control.
\newblock In \emph{Proceedings of the International Conference on Learning
  Representations}, 2023.

\bibitem[Ho et~al.(2020)Ho, Jain, and Abbeel]{ho2020denoising}
Jonathan Ho, Ajay Jain, and Pieter Abbeel.
\newblock Denoising diffusion probabilistic models.
\newblock \emph{Advances in neural information processing systems},
  33:\penalty0 6840--6851, 2020.

\bibitem[Huang et~al.(2024)Huang, Yu, Chen, Geiger, and Gao]{huang20242d}
Binbin Huang, Zehao Yu, Anpei Chen, Andreas Geiger, and Shenghua Gao.
\newblock 2d gaussian splatting for geometrically accurate radiance fields.
\newblock In \emph{ACM SIGGRAPH 2024 conference papers}, pages 1--11, 2024.

\bibitem[Huang et~al.(2026)Huang, Chng, Zhan, Yan, and Xu]{huang2026flusplat}
Haitao Huang, Shin-Fang Chng, Huangying Zhan, Qingan Yan, and Yi Xu.
\newblock Flusplat: Sparse-view 3d editing without test-time optimization.
\newblock \emph{arXiv preprint arXiv:2604.20038}, 2026.

\bibitem[Jiang et~al.(2025)Jiang, Han, Mao, Zhang, Pan, and Liu]{jiang2025vace}
Zeyinzi Jiang, Zhen Han, Chaojie Mao, Jingfeng Zhang, Yulin Pan, and Yu Liu.
\newblock Vace: All-in-one video creation and editing.
\newblock In \emph{Proceedings of the IEEE/CVF International Conference on
  Computer Vision}, pages 17191--17202, 2025.

\bibitem[Kerbl et~al.(2023)Kerbl, Kopanas, Leimk{\"u}hler, Drettakis,
  et~al.]{kerbl20233d}
Bernhard Kerbl, Georgios Kopanas, Thomas Leimk{\"u}hler, George Drettakis,
  et~al.
\newblock 3d gaussian splatting for real-time radiance field rendering.
\newblock \emph{ACM Trans. Graph.}, 42\penalty0 (4):\penalty0 139--1, 2023.

\bibitem[Ku et~al.(2024)Ku, Wei, Ren, Yang, and Chen]{ku2024anyvv}
Max Ku, Cong Wei, Weiming Ren, Huan Yang, and Wenhu Chen.
\newblock Anyv2v: A tuning-free framework for any video-to-video editing tasks.
\newblock \emph{Transactions on Machine Learning Research}, 2024.
\newblock Reproducibility Certification.

\bibitem[Labs(2024)]{flux2024}
Black~Forest Labs.
\newblock Flux.
\newblock \url{https://github.com/black-forest-labs/flux}, 2024.

\bibitem[Labs(2025)]{flux-2-2025}
Black~Forest Labs.
\newblock {FLUX.2: Frontier Visual Intelligence}.
\newblock \url{https://bfl.ai/blog/flux-2}, 2025.

\bibitem[Labs et~al.(2025)Labs, Batifol, Blattmann, Boesel, Consul, Diagne,
  Dockhorn, English, English, Esser, Kulal, Lacey, Levi, Li, Lorenz, Müller,
  Podell, Rombach, Saini, Sauer, and Smith]{labs2025flux1kontextflowmatching}
Black~Forest Labs, Stephen Batifol, Andreas Blattmann, Frederic Boesel, Saksham
  Consul, Cyril Diagne, Tim Dockhorn, Jack English, Zion English, Patrick
  Esser, Sumith Kulal, Kyle Lacey, Yam Levi, Cheng Li, Dominik Lorenz, Jonas
  Müller, Dustin Podell, Robin Rombach, Harry Saini, Axel Sauer, and Luke
  Smith.
\newblock Flux.1 kontext: Flow matching for in-context image generation and
  editing in latent space, 2025.

\bibitem[Lee et~al.(2025)Lee, Park, Seo, Park, Park, Baek, Shin, Kim, and
  Kim]{lee2025editsplat}
Dong~In Lee, Hyeongcheol Park, Jiyoung Seo, Eunbyung Park, Hyunje Park, Ha~Dam
  Baek, Sangheon Shin, Sangmin Kim, and Sangpil Kim.
\newblock Editsplat: Multi-view fusion and attention-guided optimization for
  view-consistent 3d scene editing with 3d gaussian splatting.
\newblock In \emph{Proceedings of the Computer Vision and Pattern Recognition
  Conference}, pages 11135--11145, 2025.

\bibitem[Li et~al.(2025)Li, Wu, Meng, Gao, Zhang, Wang, and
  Zhang]{li2025instancegaussian}
Haijie Li, Yanmin Wu, Jiarui Meng, Qiankun Gao, Zhiyao Zhang, Ronggang Wang,
  and Jian Zhang.
\newblock Instancegaussian: Appearance-semantic joint gaussian representation
  for 3d instance-level perception.
\newblock In \emph{Proceedings of the Computer Vision and Pattern Recognition
  Conference}, pages 14078--14088, 2025.

\bibitem[Luo et~al.(2025)Luo, Di, Yang, Ma, Xue, Chen, Gou, and
  Liu]{luo2025trame}
Chaofan Luo, Donglin Di, Xun Yang, Yongjia Ma, Zhou Xue, Wei Chen, Xiaofei Gou,
  and Yebin Liu.
\newblock Trame: Trajectory-anchored multi-view editing for text-guided 3d
  gaussian manipulation.
\newblock \emph{IEEE Transactions on Multimedia}, 2025.

\bibitem[Ma et~al.(2026)Ma, Wang, Ma, Wang, Zheng, Yang, Li, Zhao, Ying, Yang,
  et~al.]{ma2026group}
Yue Ma, Xinyu Wang, Qianli Ma, Qinghe Wang, Mingzhe Zheng, Xiangpeng Yang, Hao
  Li, Chongbo Zhao, Jixuan Ying, Harry Yang, et~al.
\newblock Group editing: Edit multiple images in one go.
\newblock In \emph{Proceedings of the IEEE/CVF Conference on Computer Vision
  and Pattern Recognition}, pages 43418--43428, 2026.

\bibitem[Meng et~al.(2022)Meng, He, Song, Song, Wu, Zhu, and
  Ermon]{meng2022sdedit}
Chenlin Meng, Yutong He, Yang Song, Jiaming Song, Jiajun Wu, Jun-Yan Zhu, and
  Stefano Ermon.
\newblock {SDE}dit: Guided image synthesis and editing with stochastic
  differential equations.
\newblock In \emph{International Conference on Learning Representations}, 2022.

\bibitem[Michel et~al.(2022)Michel, Bar-On, Liu, Benaim, and
  Hanocka]{michel2022text2mesh}
Oscar Michel, Roi Bar-On, Richard Liu, Sagie Benaim, and Rana Hanocka.
\newblock Text2mesh: Text-driven neural stylization for meshes.
\newblock In \emph{2022 IEEE/CVF Conference on Computer Vision and Pattern
  Recognition (CVPR)}, pages 13482--13492. IEEE, 2022.

\bibitem[Mokady et~al.(2023)Mokady, Hertz, Aberman, Pritch, and
  Cohen-Or]{mokady2023null}
Ron Mokady, Amir Hertz, Kfir Aberman, Yael Pritch, and Daniel Cohen-Or.
\newblock Null-text inversion for editing real images using guided diffusion
  models.
\newblock In \emph{Proceedings of the IEEE/CVF conference on computer vision
  and pattern recognition}, pages 6038--6047, 2023.

\bibitem[Palandra et~al.(2024)Palandra, Sanchietti, Baieri, and
  Rodola]{palandra2024gsedit}
Francesco Palandra, Andrea Sanchietti, Daniele Baieri, and Emanuele Rodola.
\newblock Gsedit: Efficient text-guided editing of 3d objects via gaussian
  splatting.
\newblock \emph{arXiv preprint arXiv:2403.05154}, 2024.

\bibitem[Qu et~al.(2025)Qu, Zhang, Luo, Qi, Zhang, Liu, Sengupta, and
  Chen]{qu2025editcast3d}
Huaizhi Qu, Ruichen Zhang, Shuqing Luo, Luchao Qi, Zhihao Zhang, Xiaoming Liu,
  Roni Sengupta, and Tianlong Chen.
\newblock Editcast3d: Single-frame-guided 3d editing with video propagation and
  view selection.
\newblock \emph{arXiv preprint arXiv:2510.13652}, 2025.

\bibitem[Rombach et~al.(2022)Rombach, Blattmann, Lorenz, Esser, and
  Ommer]{rombach2022high}
Robin Rombach, Andreas Blattmann, Dominik Lorenz, Patrick Esser, and Bj{\"o}rn
  Ommer.
\newblock High-resolution image synthesis with latent diffusion models.
\newblock In \emph{Proceedings of the IEEE/CVF conference on computer vision
  and pattern recognition}, pages 10684--10695, 2022.

\bibitem[Shi et~al.(2024)Shi, Huang, Wang, Bian, Li, Zhang, Zhang, Cheung, See,
  Qin, et~al.]{shi2024motion}
Xiaoyu Shi, Zhaoyang Huang, Fu-Yun Wang, Weikang Bian, Dasong Li, Yi Zhang,
  Manyuan Zhang, Ka~Chun Cheung, Simon See, Hongwei Qin, et~al.
\newblock Motion-i2v: Consistent and controllable image-to-video generation
  with explicit motion modeling.
\newblock In \emph{ACM SIGGRAPH 2024 Conference Papers}, pages 1--11, 2024.

\bibitem[Song et~al.(2021)Song, Meng, and Ermon]{song2021denoising}
Jiaming Song, Chenlin Meng, and Stefano Ermon.
\newblock Denoising diffusion implicit models.
\newblock In \emph{International Conference on Learning Representations}, 2021.

\bibitem[Wang et~al.(2024)Wang, Yi, Wu, Zhao, Chen, and Zhang]{wang2024view}
Yuxuan Wang, Xuanyu Yi, Zike Wu, Na Zhao, Long Chen, and Hanwang Zhang.
\newblock View-consistent 3d editing with gaussian splatting.
\newblock In \emph{European conference on computer vision}, pages 404--420.
  Springer, 2024.

\bibitem[Wen et~al.(2025)Wen, Wu, Wang, and Liang]{wen2025intergsedit}
Minghao Wen, Shengjie Wu, Kangkan Wang, and Dong Liang.
\newblock Intergsedit: Interactive 3d gaussian splatting editing with 3d
  geometry-consistent attention prior.
\newblock In \emph{Proceedings of the IEEE/CVF International Conference on
  Computer Vision}, pages 26136--26145, 2025.

\bibitem[Wu et~al.(2025)Wu, Li, Zhou, Lin, Gao, Yan, Yin, Bai, Xu, Chen,
  et~al.]{wu2025qwen}
Chenfei Wu, Jiahao Li, Jingren Zhou, Junyang Lin, Kaiyuan Gao, Kun Yan,
  Sheng-ming Yin, Shuai Bai, Xiao Xu, Yilei Chen, et~al.
\newblock Qwen-image technical report.
\newblock \emph{arXiv preprint arXiv:2508.02324}, 2025.

\bibitem[Wu et~al.(2024)Wu, Bian, Li, Wang, Reid, Torr, and
  Prisacariu]{wu2024gaussctrl}
Jing Wu, Jia-Wang Bian, Xinghui Li, Guangrun Wang, Ian~D. Reid, Philip Torr,
  and Victor~Adrian Prisacariu.
\newblock Gaussctrl: Multi-view consistent text-driven 3d gaussian splatting
  editing.
\newblock In \emph{Proceedings of the European Conference on Computer Vision},
  pages 55--71. Springer, 2024.

\bibitem[Xiao et~al.(2025)Xiao, Chen, Huang, Xiong, Yang, Prasad, and
  Zhao]{xiao2025localized}
Hanyuan Xiao, Yingshu Chen, Huajian Huang, Haolin Xiong, Jing Yang, Pratusha
  Prasad, and Yajie Zhao.
\newblock Localized gaussian splatting editing with contextual awareness.
\newblock In \emph{2025 IEEE/CVF Winter Conference on Applications of Computer
  Vision (WACV)}, pages 5207--5217. IEEE, 2025.

\bibitem[Xu et~al.(2024)Xu, Huang, Pan, Ma, and Chai]{xu2024inversionfree}
Sihan Xu, Yidong Huang, Jiayi Pan, Ziqiao Ma, and Joyce Chai.
\newblock Inversion-free image editing with language-guided diffusion models.
\newblock In \emph{Proceedings of the IEEE/CVF Conference on Computer Vision
  and Pattern Recognition}, pages 9454--9461, 2024.

\bibitem[Zhang et~al.(2025)Zhang, Jiang, Chen, Jiang, and Liu]{zhang2025cob}
Jiaxin Zhang, Junjun Jiang, Youyu Chen, Kui Jiang, and Xianming Liu.
\newblock Cob-gs: Clear object boundaries in 3dgs segmentation based on
  boundary-adaptive gaussian splitting.
\newblock In \emph{Proceedings of the IEEE/CVF Conference on Computer Vision
  and Pattern Recognition}, pages 19335--19344, 2025.

\bibitem[Zhang et~al.(2023)Zhang, Rao, and Agrawala]{zhang2023adding}
Lvmin Zhang, Anyi Rao, and Maneesh Agrawala.
\newblock Adding conditional control to text-to-image diffusion models.
\newblock In \emph{Proceedings of the IEEE/CVF international conference on
  computer vision}, pages 3836--3847, 2023.

\bibitem[Zhao et~al.(2026)Zhao, Wu, Li, Meng, Li, Zhang, Zhou, Lin, Gao, Cao,
  et~al.]{zhao2026qwen}
Bing Zhao, Chenfei Wu, Deqing Li, Hao Meng, Jiahao Li, Jie Zhang, Jingren Zhou,
  Junyang Lin, Kaiyuan Gao, Kuan Cao, et~al.
\newblock Qwen-image-2.0 technical report.
\newblock \emph{arXiv preprint arXiv:2605.10730}, 2026.

\bibitem[Zhu et~al.(2026)Zhu, Chen, Li, and Wei]{zhu2026coreeditor}
Zhe Zhu, Honghua Chen, Peng Li, and Mingqiang Wei.
\newblock Coreeditor: Correspondence-constrained diffusion for consistent 3d
  editing.
\newblock \emph{IEEE Transactions on Visualization and Computer Graphics},
  32\penalty0 (3):\penalty0 2838--2851, 2026.

\bibitem[Zhuang et~al.(2023)Zhuang, Wang, Lin, Liu, and
  Li]{zhuang2023dreameditor}
Jingyu Zhuang, Chen Wang, Liang Lin, Lingjie Liu, and Guanbin Li.
\newblock Dreameditor: Text-driven 3d scene editing with neural fields.
\newblock In \emph{SIGGRAPH Asia 2023 conference papers}, pages 1--10, 2023.

\bibitem[Zhuang et~al.(2024)Zhuang, Kang, Cao, Li, Lin, and
  Shan]{zhuang2024tip}
Jingyu Zhuang, Di Kang, Yan-Pei Cao, Guanbin Li, Liang Lin, and Ying Shan.
\newblock Tip-editor: An accurate 3d editor following both text-prompts and
  image-prompts.
\newblock \emph{ACM Transactions on Graphics (ToG)}, 43\penalty0 (4):\penalty0
  1--12, 2024.

\end{thebibliography}
}

\clearpage
\maketitlesupplementary
\appendix
\setcounter{figure}{0}
\setcounter{table}{0}
\setcounter{equation}{0}
\setcounter{algorithm}{0}
\renewcommand{\thefigure}{S\arabic{figure}}
\renewcommand{\thetable}{S\arabic{table}}
\renewcommand{\theequation}{S\arabic{equation}}
\renewcommand{\thealgorithm}{S\arabic{algorithm}}
\renewcommand*{\theHfigure}{supp.\arabic{figure}}
\renewcommand*{\theHtable}{supp.\arabic{table}}
\renewcommand*{\theHequation}{supp.\arabic{equation}}
\renewcommand*{\theHalgorithm}{supp.\arabic{algorithm}}
\setcounter{topnumber}{1}

\newcolumntype{L}[1]{>{\raggedright\arraybackslash}m{#1}}
\newcolumntype{K}[1]{>{\centering\arraybackslash}m{#1}}
\newcommand{\taskref}[1]{\begin{minipage}[c][0.86in][c]{\linewidth}\centering\includegraphics[width=0.96\linewidth,height=0.76in,keepaspectratio]{#1}\end{minipage}}

This supplementary material provides implementation and evaluation details, additional analyses of editing supervision, further qualitative results, and the instructions and reference images used for the twelve quantitative editing tasks.

\section{Implementation and Evaluation Details}
\label{sec:supp-details}

\subsection{Implementation Details}

Algorithm~\ref{alg:edit2meshgs} summarizes Gaussian tagging, reference-guided optimization, residual pruning, and mesh extraction.

\textbf{Optimization Settings.}
All images are edited at a resolution of $512\times512$. We use 5,000 iterations for large deformation edits and 4,000 iterations for the remaining tasks. A set of 20 edited supervision views is regenerated every 1,000 iterations. The first 2,000 iterations use only the appearance losses $\mathcal{L}_{1}$ and $\mathcal{L}_{p}$, while the remaining iterations use the full objective. We set $\lambda_1=0.8$, $\lambda_p=0.8$, $\lambda_d=0.05$, $\lambda_n=0.05$, and $\lambda_s=0.05$. To mitigate local rendering artifacts that STP may introduce, we optionally refine the remaining Gaussians for an additional 1,000 iterations before mesh extraction.

\textbf{SVGT Settings.}
Initial selection and verification each use 20 randomly sampled views, with no overlap between the two sets. The default candidate retention ratio is $\tau=0.7$, the veto weight is $\lambda_{\mathrm{veto}}=5$, and the final tagging threshold is $\theta_{\mathrm{tag}}=0.7$.

\textbf{STP Settings.}
After editing, STP uses the final tangent-scale product $a_i=s_i^{u}(T)s_i^{v}(T)$ as the geometric weight. We rank the Gaussians in the final editable set $\mathcal{G}_{\mathrm{edit}}^{T}$ by $r_i=s_i(1-u_i)a_i$ in descending order and prune the top 30\%, setting $\tau_{\mathrm p}=0.3$. STP is applied when both source and target prompts are provided. The remaining Gaussians are briefly refined before mesh extraction.

\textbf{Comparison Settings.}
For GaussianEditor$^{*}$, we directly replace its editing model with FLUX.2-klein-9B while retaining its official default parameters and single-view iterative editing scheme. In contrast, the 3DGS + FLUX.2 configuration in our ablation study follows the same multi-view iterative editing scheme as Edit2MeshGS. Thus, GaussianEditor$^{*}$ and 3DGS + FLUX.2 are distinct configurations despite sharing the same editing model. For Omni-3DEdit$^{*}$, we replace its original single-image editing model with FLUX.2-klein-9B while retaining the remaining official optimization settings. GaussianEditor$^{*}$, Omni-3DEdit$^{*}$, and Edit2MeshGS use the same editing instructions and reference images. For Gaussian segmentation, all compared methods use 2D masks generated by SAM3.

\textbf{Mesh Extraction.}
Following 2DGS, we render depth maps from the training views and fuse them with TSDF using Open3D. The voxel size is $0.004$, and the truncation threshold is $0.02$. Meshes from the other 3DGS-based methods are extracted in the same manner with these settings.

\begin{algorithm}[!t]
\footnotesize
\algrenewcommand\algorithmicindent{1em}
\caption{Edit2MeshGS Pipeline}
\label{alg:edit2meshgs}
\begin{algorithmic}[1]
\raggedright
\Require Initial 2DGS scene $\mathcal{G}^{0}$, camera views $\mathcal{V}$,
editing instruction $P$, optional style reference image $R_{\mathrm{img}}$,
optional source prompt $q_s$, optional target prompt $q_t$, maximum iterations $T$
\Ensure Edited 2DGS scene $\widetilde{\mathcal{G}}$ and reconstructed mesh $\mathcal{M}$

\State Render initial views $\mathcal{I}_{\mathrm{render}}^{0}$ from $\mathcal{G}^{0}$
\If{$q_s \neq \emptyset$}
    \State Obtain source masks $\mathbf{M}_{s}^{0} \gets \mathrm{SAM3}(\mathcal{I}_{\mathrm{render}}^{0}, q_s)$
    \State Obtain editable Gaussian set $\mathcal{G}_{\mathrm{edit}}^{0} \gets \mathrm{SVGT}(\mathcal{G}^{0}, \mathbf{M}_{s}^{0}, \mathcal{V})$
    \State Compute STP source association scores $\{s_i\} \gets \mathrm{Assoc}(\mathcal{G}^{0}, \mathbf{M}_{s}^{0}, \mathcal{V})$
\Else
    \State Set $\mathcal{G}_{\mathrm{edit}}^{0} \gets \mathcal{G}^{0}$
\EndIf
\State Define editable tags $m_i \gets \mathbf{1}[g_i^0\in\mathcal{G}_{\mathrm{edit}}^0]$

\For{$t=0$ to $T-1$}
    \If{$t \bmod 1000 = 0$}
        \State Randomly sample 20 views $\mathcal{V}^{t}\subset\mathcal{V}$
        \State Render $\mathcal{I}_{\mathrm{render}}^{t}$ from $\mathcal{G}^{t}$ over $\mathcal{V}^{t}$
        \State Generate edited views
        $\mathcal{I}_{\mathrm{edit}}^{t} \gets \Phi_{\mathrm{FLUX2}}(\mathcal{I}_{\mathrm{render}}^{t}, R_{\mathrm{img}}, P)$
    \EndIf
    \State Set $\mathcal{G}_{\mathrm{edit}}^{t} \gets \{g_i^t\in\mathcal{G}^{t}\mid m_i=1\}$
    \State Optimize only $\mathcal{G}_{\mathrm{edit}}^{t}$ while fixing background Gaussians
    \If{$t < 2000$}
        \State Use appearance loss $\lambda_1\mathcal{L}_{1}+\lambda_p\mathcal{L}_{p}$
    \Else
        \State Use full loss $\lambda_1\mathcal{L}_{1}+\lambda_p\mathcal{L}_{p}+\lambda_d\mathcal{L}_{d}+\lambda_n\mathcal{L}_{n}+\lambda_s\mathcal{L}_{s}$
    \EndIf
    \State Update $\mathcal{G}^{t}$ to $\mathcal{G}^{t+1}$ by gradient descent
\EndFor

\State Set $\mathcal{G}^{\mathrm{final}} \gets \mathcal{G}^{T}$ and $\mathcal{G}_{\mathrm{edit}}^{T} \gets \{g_i\in\mathcal{G}^{\mathrm{final}}\mid m_i=1\}$
\If{$q_s \neq \emptyset$ and $q_t \neq \emptyset$}
    \State Render all final views $\mathcal{I}_{\mathrm{render}}^{\mathrm{final}}$ from $\mathcal{G}^{\mathrm{final}}$
    \State Obtain target masks $\mathbf{M}_{t} \gets \mathrm{SAM3}(\mathcal{I}_{\mathrm{render}}^{\mathrm{final}}, q_t)$
    \State Compute target association scores $\{u_i\} \gets \mathrm{Assoc}(\mathcal{G}^{\mathrm{final}}, \mathbf{M}_{t}, \mathcal{V})$
    \State Compute geometric weights $a_i \gets s_i^{u}(T)s_i^{v}(T)$ for $g_i\in\mathcal{G}_{\mathrm{edit}}^{T}$
    \State Compute residual scores $r_i \gets s_i(1-u_i)a_i$
    \State Select the top $\tau_{\mathrm p}$ fraction by $r_i$: $\mathcal{G}_{\mathrm{res}} \gets \operatorname{Top}_{\tau_{\mathrm p}}(\mathcal{G}_{\mathrm{edit}}^{T};r)$
    \State Prune $\widetilde{\mathcal{G}} \gets \mathcal{G}^{\mathrm{final}}\setminus\mathcal{G}_{\mathrm{res}}$, then briefly refine it to reduce local discontinuities
\Else
    \State Set $\widetilde{\mathcal{G}} \gets \mathcal{G}^{\mathrm{final}}$
\EndIf
\State Render depth maps $\mathcal{D}_{\mathrm{img}}$ from $\widetilde{\mathcal{G}}$ over training views
\State Reconstruct mesh $\mathcal{M} \gets \mathrm{TSDF}(\mathcal{D}_{\mathrm{img}})$
\State \textbf{return} $\widetilde{\mathcal{G}}, \mathcal{M}$
\end{algorithmic}
\end{algorithm}

\setcounter{dbltopnumber}{1}
\begin{figure*}[t]
\centering
    \includegraphics[width=0.97\textwidth]{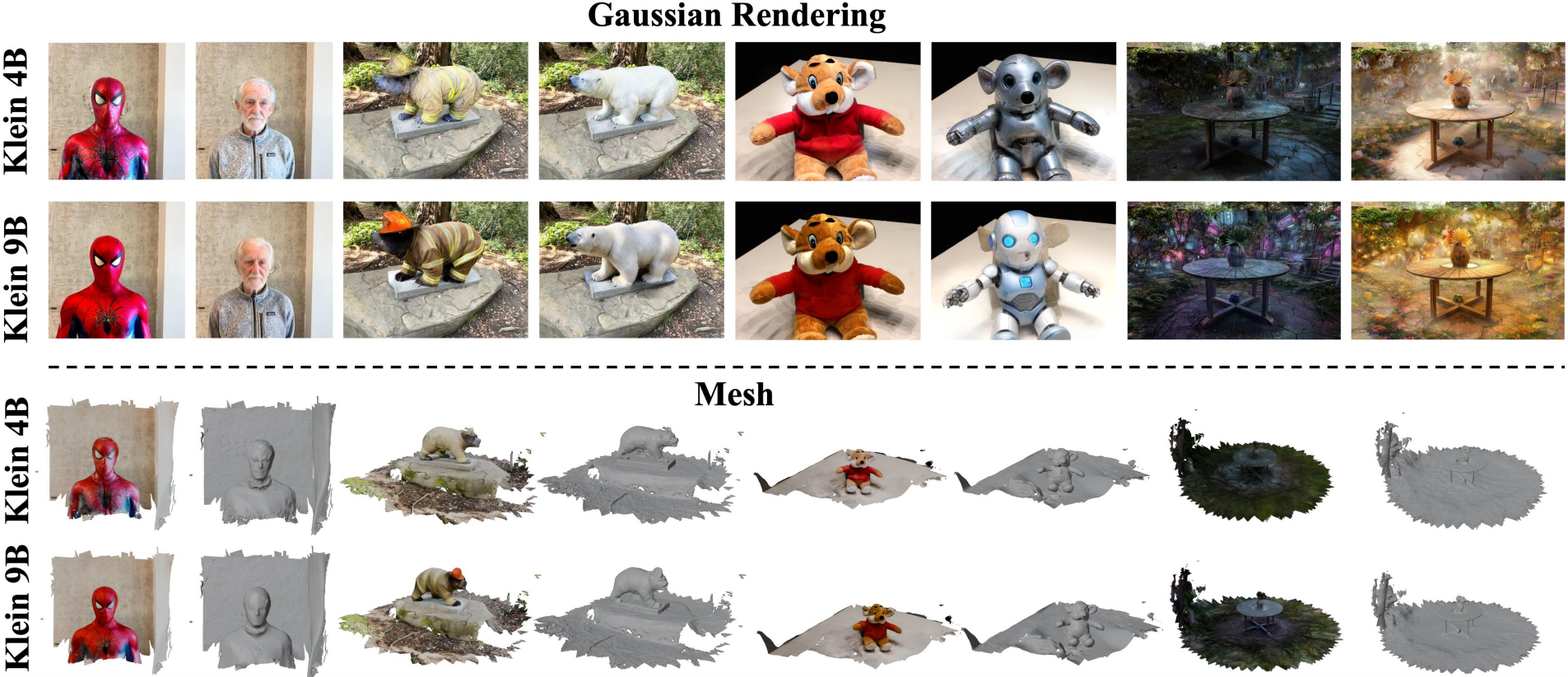}
    \caption{Qualitative comparison between FLUX.2-klein-4B and FLUX.2-klein-9B in Edit2MeshGS. The figure shows edited renderings and reconstructed meshes for representative tasks included in the quantitative comparison.}
    \label{fig:flux-scale}
\end{figure*}

\begin{figure}[t]
    \centering
    \includegraphics[width=0.98\columnwidth]{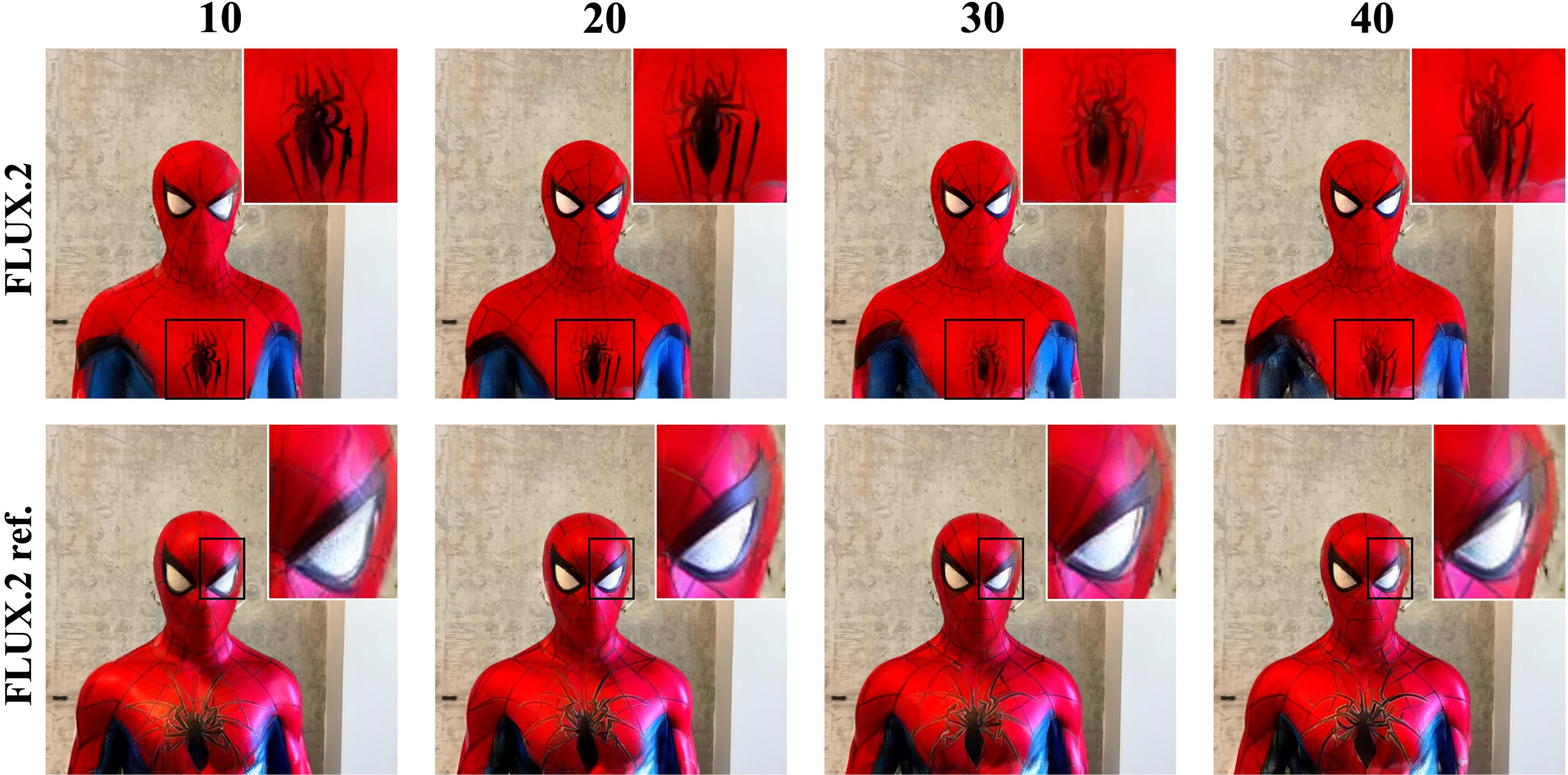}
    \caption{Effect of using 10, 20, 30, and 40 edited supervision views. Close-ups highlight chest-pattern and eye-boundary details with and without shared reference guidance.}
    \label{fig:numview}
\end{figure}

\subsection{Evaluation Protocols}

\textbf{Rendering Metrics.}
We report CLIP Directional Similarity (CDS), CLIP Similarity (CS), and 3D Consistency (3DC). CDS measures alignment between image and text editing directions, CS measures target semantic alignment, and 3DC evaluates cross-view consistency of the editing directions. These metrics are averaged over the twelve editing tasks listed in Sec.~\ref{sec:supp-tasks}.

\textbf{3D Consistency.}
For an editing task with $N\geq2$ evaluation views, let $I_s^v$ and $I_e^v$ denote the original and edited scene renderings from the same camera at view $v$. We extract image features using CLIP ViT-B/32 with its standard preprocessing, including resizing the shorter side to 224 pixels and center-cropping to $224\times224$. The normalized feature is
\begin{equation}
g(I)=\frac{f_I(I)}{\max(\|f_I(I)\|_2,\epsilon)},
\end{equation}
where $f_I$ is the image encoder and $\epsilon=10^{-8}$ provides numerical stability. We then compute the feature change and its normalized direction:
\begin{equation}
\Delta_v=g(I_e^v)-g(I_s^v),\qquad
 d_v=\frac{\Delta_v}{\max(\|\Delta_v\|_2,\epsilon)}.
\end{equation}
The difference $\Delta_v$ represents the semantic change induced by editing at view $v$. Normalization makes the comparison depend on the direction rather than the magnitude of this change. The task-level consistency is
\begin{equation}
C=\frac{2}{N(N-1)}\sum_{1\leq i<j\leq N} d_i^{\top}d_j,
\label{eq:supp-3dc}
\end{equation}
which averages over all $N(N-1)/2$ unordered pairs of distinct views. For nondegenerate directions, each dot product is a cosine similarity. Zero feature changes remain zero and contribute zero to the pairwise scores; near-zero changes use the same $\epsilon$ guard. We report $\mathrm{3DC}=\frac{100}{K}\sum_{k=1}^{K}C_k$ over the $K=12$ editing tasks, giving each task equal weight. Higher scores indicate more consistent semantic editing directions across views.

\textbf{Mesh Metrics.}
We evaluate the extracted meshes using Small Component Area Ratio (SCAR), Normal Variation (NV), and Boundary Edge Ratio (BER), averaged over the same twelve editing tasks. SCAR and BER are ratios, whereas NV is reported in degrees. Their definitions are given below.

\textbf{Small Component Area Ratio.}
SCAR measures the surface area ratio of small connected components:
\begin{equation}
\mathrm{SCAR}=
\frac{\sum_{c\in\mathcal{C}_{s}} A(c)}{A(\mathcal{M})},
\end{equation}
where $A(\mathcal{M})$ is the total mesh surface area and $\mathcal{C}_{s}$ denotes connected components whose area is smaller than $\tau_{a}A(\mathcal{M})$. We build connected components by face adjacency through shared edges, and set $\tau_{a}=0.01$ in all experiments. A lower SCAR indicates a smaller fraction of surface area in small connected components.

\textbf{Normal Variation.}
NV measures local surface roughness by averaging the angle difference between adjacent face normals:
\begin{equation}
\mathrm{NV}=
\frac{180}{\pi|\mathcal{E}_{m}|}
\sum_{(f_i,f_j)\in\mathcal{E}_{m}}
\arccos\left(|\mathbf{n}_{i}^{\top}\mathbf{n}_{j}|\right),
\end{equation}
where $\mathbf{n}_{i}$ and $\mathbf{n}_{j}$ are unit face normals, and $\mathcal{E}_{m}$ contains manifold adjacent face pairs sharing one edge. Degenerate zero area faces are ignored. We use the absolute dot product to reduce the influence of inconsistent face winding. NV is reported in degrees, and lower values indicate smoother local surfaces.

\begin{figure}[t]
    \centering
    \includegraphics[width=0.98\columnwidth]{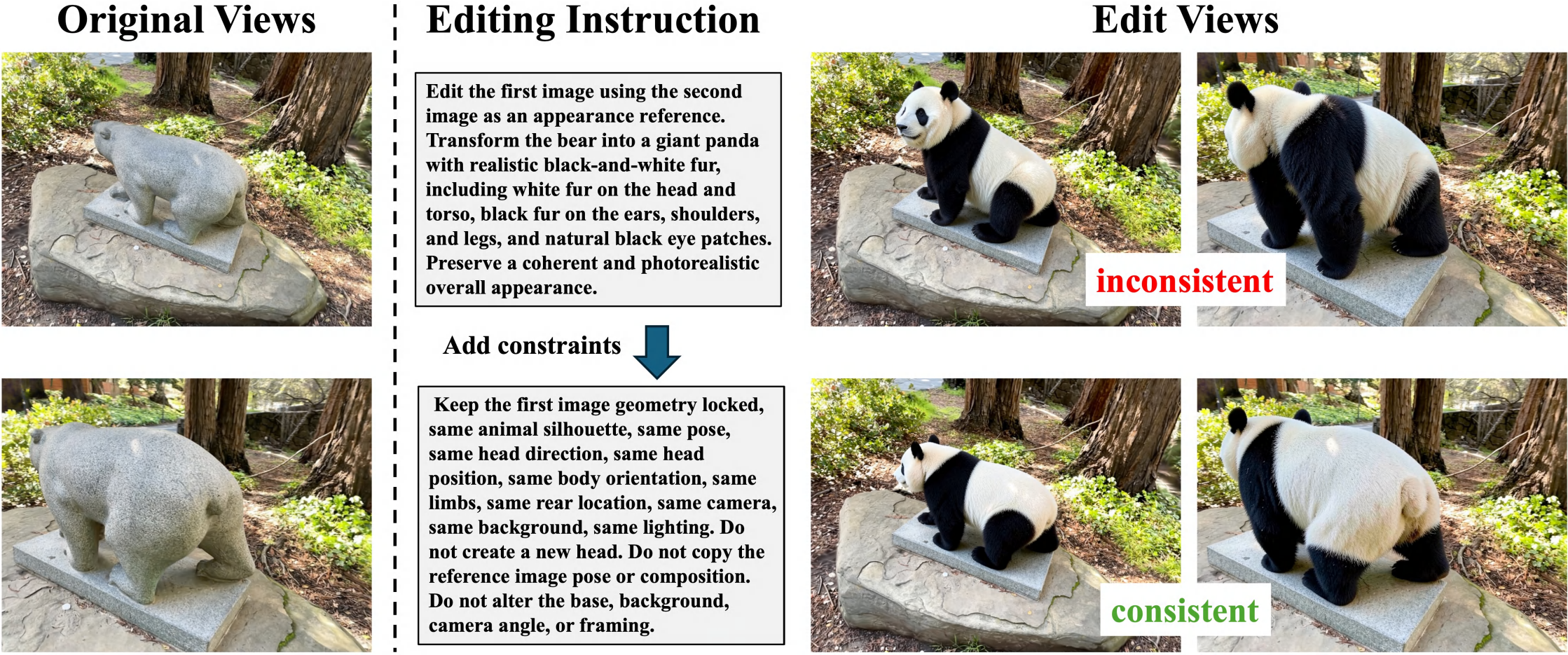}
    \caption{Effect of structural constraints in editing instructions. A generic appearance prompt produces inconsistent structures across views, while explicit pose and geometry constraints improve cross view consistency.}
    \label{fig:instruction}
\end{figure}

\begin{figure}[t]
    \centering
    \includegraphics[width=0.98\columnwidth]{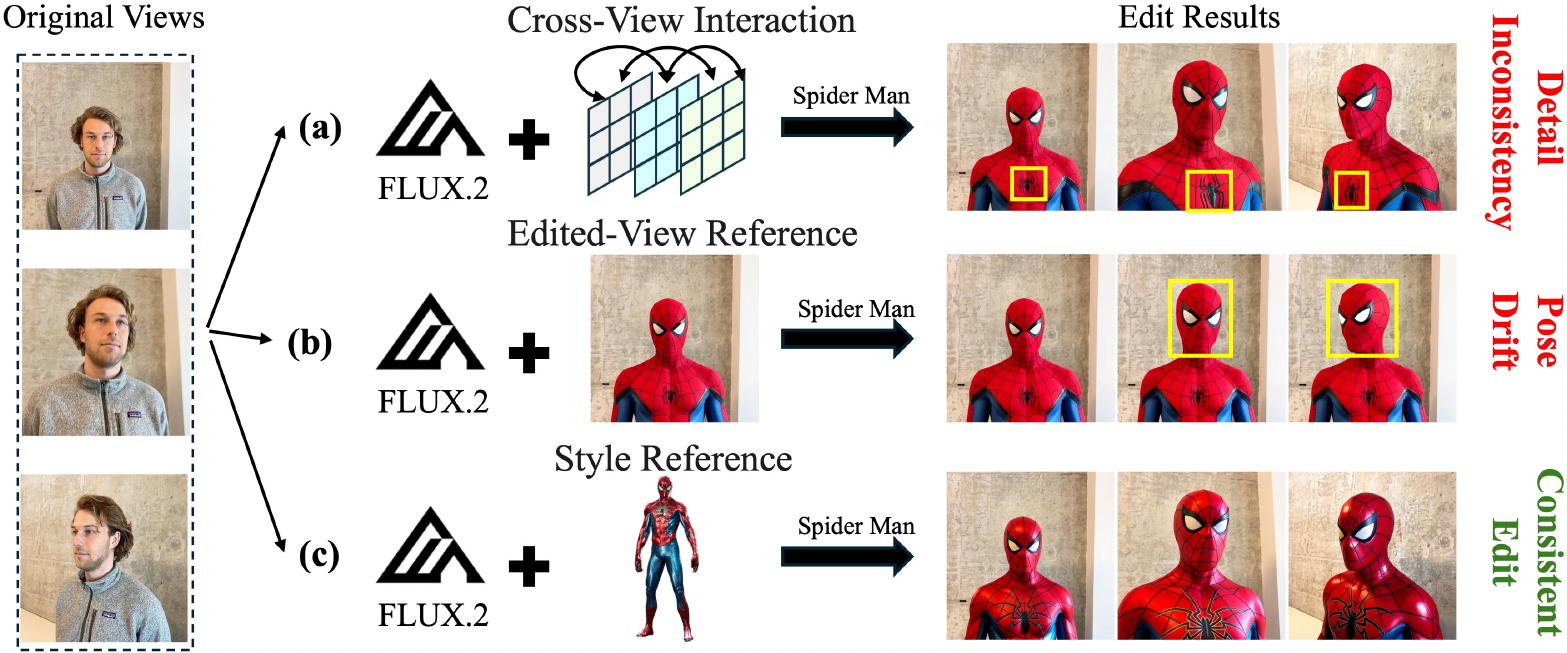}
    \caption{Comparison of multi-view editing strategies. Cross-view interaction lacks fine-grained details, while conditioning on a single edited view may confuse the model. A style reference provides clearer guidance and achieves consistent edits across views.}
    \label{fig:flux-comp}
\end{figure}

\textbf{Boundary Edge Ratio.}
BER measures the ratio of open boundary edges:
\begin{equation}
\mathrm{BER}=
\frac{|\mathcal{E}_{b}|}{|\mathcal{E}|},
\end{equation}
where $\mathcal{E}$ is the set of unique mesh edges and $\mathcal{E}_{b}$ contains edges used by exactly one triangle. Under the same scene and extraction setting, a lower BER indicates fewer holes, open boundaries, and discontinuities.

\begin{figure*}[!t]
    \centering
    \includegraphics[width=0.98\textwidth]{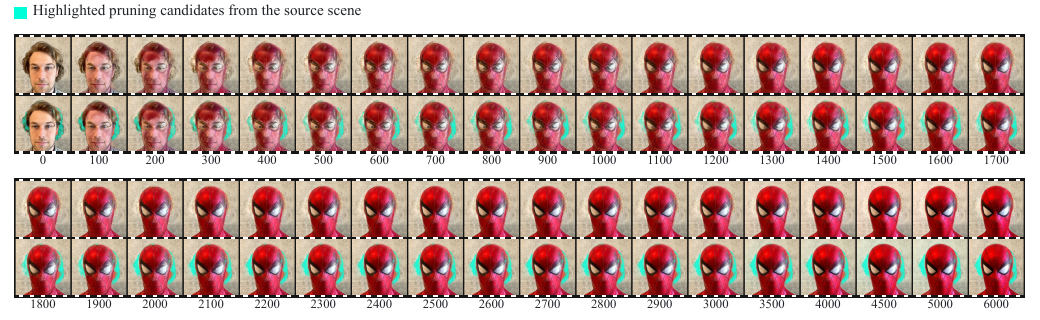}
    \caption{Appearance changes and projected pruning candidates during the Spider-Man edit. For each checkpoint, the top row shows the RGB rendering from a fixed view. The bottom row overlays the projected STP candidates on that rendering. Teal marks candidates from the original hair region and their descendants. Even after the edited head takes on the appearance of the Spider-Man mask, Gaussians from the original hair region still contribute to its outline.}
    \label{fig:supp-hair-rgb}
\end{figure*}

\begin{figure*}[!t]
    \centering
    \includegraphics[width=0.98\textwidth]{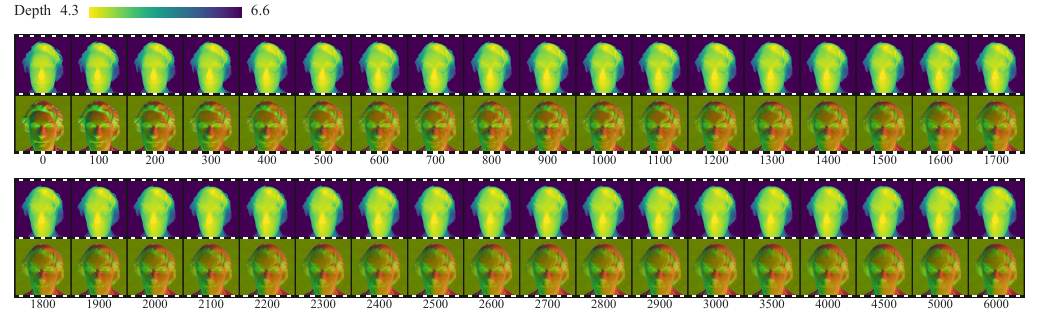}
    \caption{Depth and surface normals during the Spider-Man edit. Each checkpoint shows expected depth (top) and camera-space surface normals (bottom) for the same view and crop as Fig.~\ref{fig:supp-hair-rgb}. All depth maps use the same scale, from 4.3 (near, yellow) to 6.6 (far, purple) in scene units. The small median change in depth within the original hair region is consistent with partial retention of its surface shape.}
    \label{fig:supp-hair-depth}
\end{figure*}

\begin{figure*}[!t]
    \centering
    \includegraphics[width=0.98\textwidth]{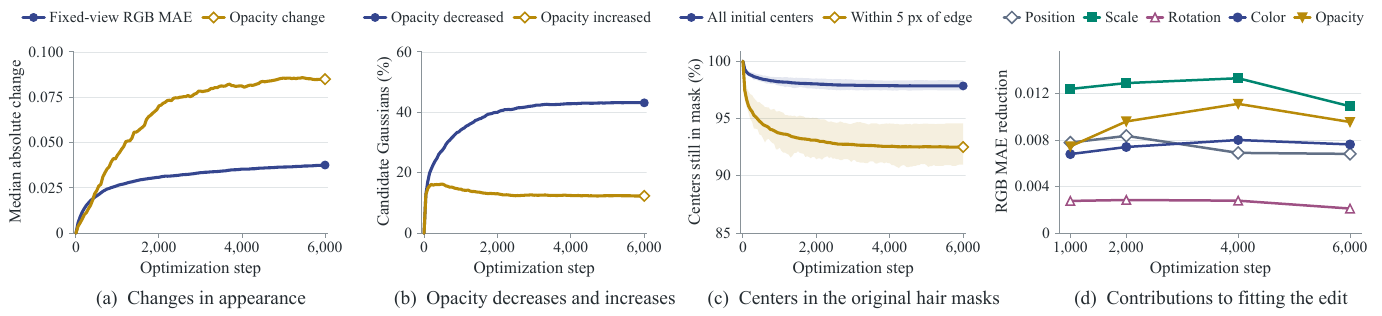}
    \caption{Appearance changes and projected centers of Gaussians from the original hair region. (a) Median absolute changes in color and opacity from initialization. (b) Fractions of Gaussians whose opacity increases or decreases by more than 0.05. (c) Fraction of centers initially projected inside each view's hair mask that still project inside it. Lines show means across 65 views, and shaded bands span the 25th to 75th percentiles. (d) Contributions of position, scale, rotation, color, and opacity to the reduction in RGB mean absolute error within the hair masks.}
    \label{fig:supp-hair-curves}
\end{figure*}

\textbf{Segmentation Evaluation.}
The evaluation pseudo-labels are SAM3 segmentation masks generated with additional point prompts. We project the tagged Gaussian masks onto the rendered views and compute precision, recall, and foreground IoU for each view, then average all three metrics over the same views with equal weights. Gaussian tagging uses 40 randomly sampled views, whereas evaluation uses the complete view set, including views not selected for tagging. We report results over five local object cases from the Face, Stone Bear, Doll, Garden, and Bonsai scenes.

\textbf{User Study.}
We conduct a user study with 60 participants to evaluate the rendering fidelity and mesh quality of the edited results separately. The two US columns in the main paper report participants' preferences as percentages for rendered views and reconstructed meshes, respectively.

\textbf{Efficiency.}
Editing time and GPU memory usage are reported on a single NVIDIA L20 GPU. Segmentation efficiency is averaged over five scenes. Tag Time denotes query time for the compared methods or SVGT tagging time using pre-segmented results.

\section{Additional Analyses}
\label{sec:supp-analyses}

\subsection{Evidence of Residual Source Structures}
\label{sec:supp-source-structures}

To examine whether the edited head retains any of the original hair structure, we use STP at step 2,000 to identify pruning candidates inside the previous editable region but outside the updated target region. We keep these candidates in the scene and continue optimization through step 6,000. As shown in Fig.~\ref{fig:supp-hair-rgb}, candidates from the original hair region continue to contribute to the rendered outline of the red mask as it takes shape. Fig.~\ref{fig:supp-hair-depth} shows how the rendered depth and surface normals change in the same view. Within the original hair mask, the median absolute relative change in depth from initialization to step 6,000 is 1.72\%, and the median angle between the initial and final surface normals is $18.14^\circ$. These observations suggest that some of the original surface geometry persists in the hair region even after the Spider-Man mask becomes visible in the rendering.

Fig.~\ref{fig:supp-hair-curves} shows how the appearance and projected centers of candidates from the original hair region change during editing. At step 6,000, 97.84\% of the centers that initially projected inside the hair mask still project inside it, on average across 65 views. For the subset initially within five pixels of the mask boundary, this fraction is 92.51\%. To measure how these Gaussians contribute to fitting the edited appearance, we hold the other Gaussians fixed and compare rendering errors across combinations of initial and edited values for the candidates' parameters. Fig.~\ref{fig:supp-hair-curves}d shows that, across the supervised views at step 6,000, changes in color and opacity together account for 46.4\% of the reduction in RGB mean absolute error within the hair masks. Together, these visual and quantitative results support the observation in the main paper that parts of the source structure can persist even after the edited scene takes on the target appearance.

\subsection{Effect of the Number of Edited Views}

Fig.~\ref{fig:numview} compares 10, 20, 30, and 40 edited supervision views with and without a shared style reference. Without a reference, fine details such as the chest emblem vary across edited views. Optimizing against more views can average out these inconsistent details and blur the emblem. With a shared reference, the edits are more consistent, and the eye boundary is better defined with 20 views than with 10. Increasing the count to 30 or 40 can still introduce small artifacts from individual views. In this comparison, 20 views give the best balance between preserving detail and limiting artifacts in both settings. We therefore use 20 edited views at each supervision refresh.

\subsection{Structural Constraints in Editing Instructions}

Fig.~\ref{fig:instruction} compares an appearance instruction with a version augmented by pose and geometry constraints. These constraints help preserve consistent object pose and structure across the illustrated views. The task-specific instructions are provided in Sec.~\ref{sec:supp-tasks}.

\section{Additional Qualitative Results}
\label{sec:supp-qualitative}

\begin{figure*}[p]
    \centering
    \includegraphics[width=\textwidth,height=0.92\textheight,keepaspectratio]{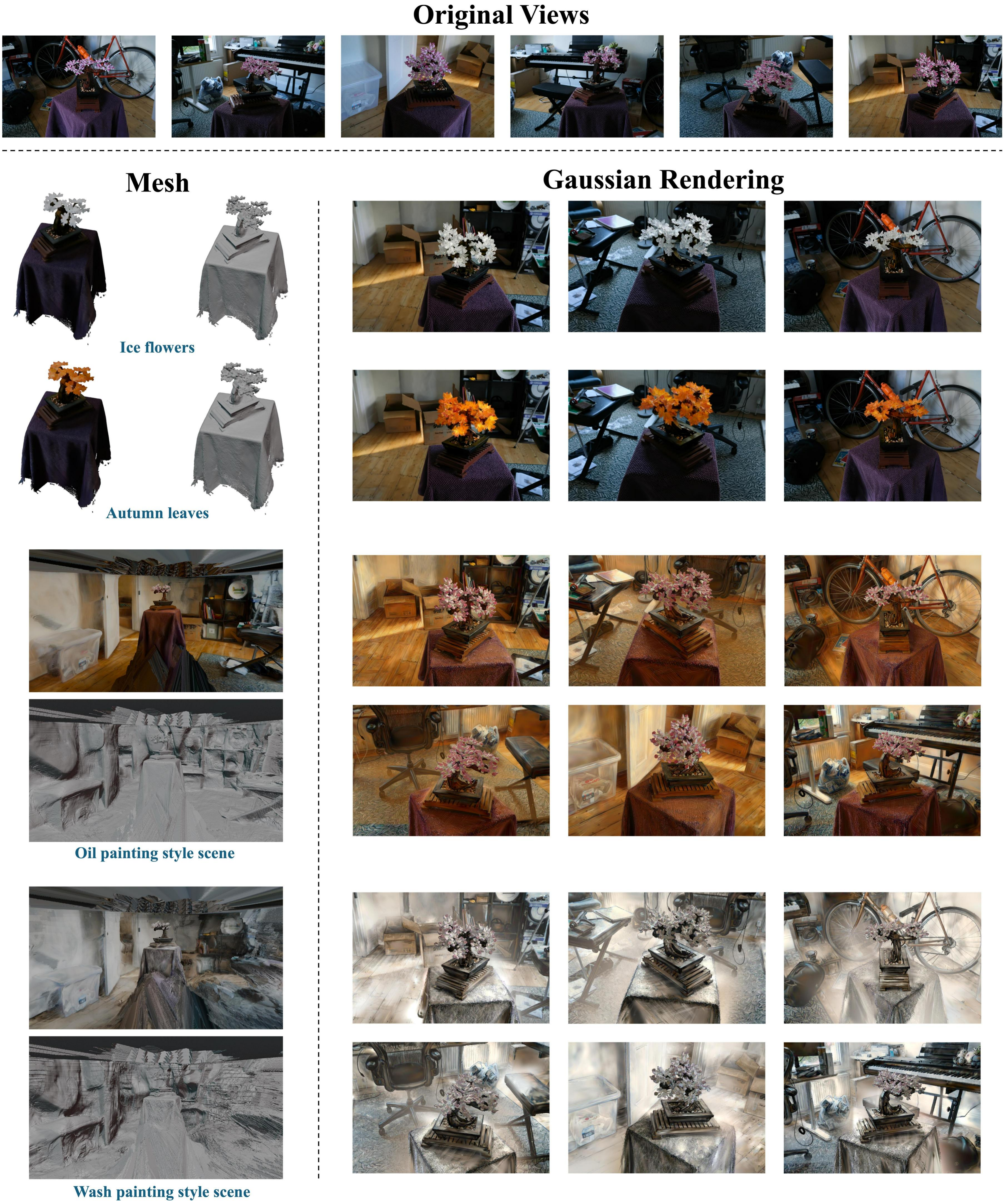}
    \caption{Additional qualitative result of Edit2MeshGS.}
    \label{fig:supp-add-zs-1}
\end{figure*}

\begin{figure*}[p]
    \centering
    \includegraphics[width=\textwidth,height=0.92\textheight,keepaspectratio]{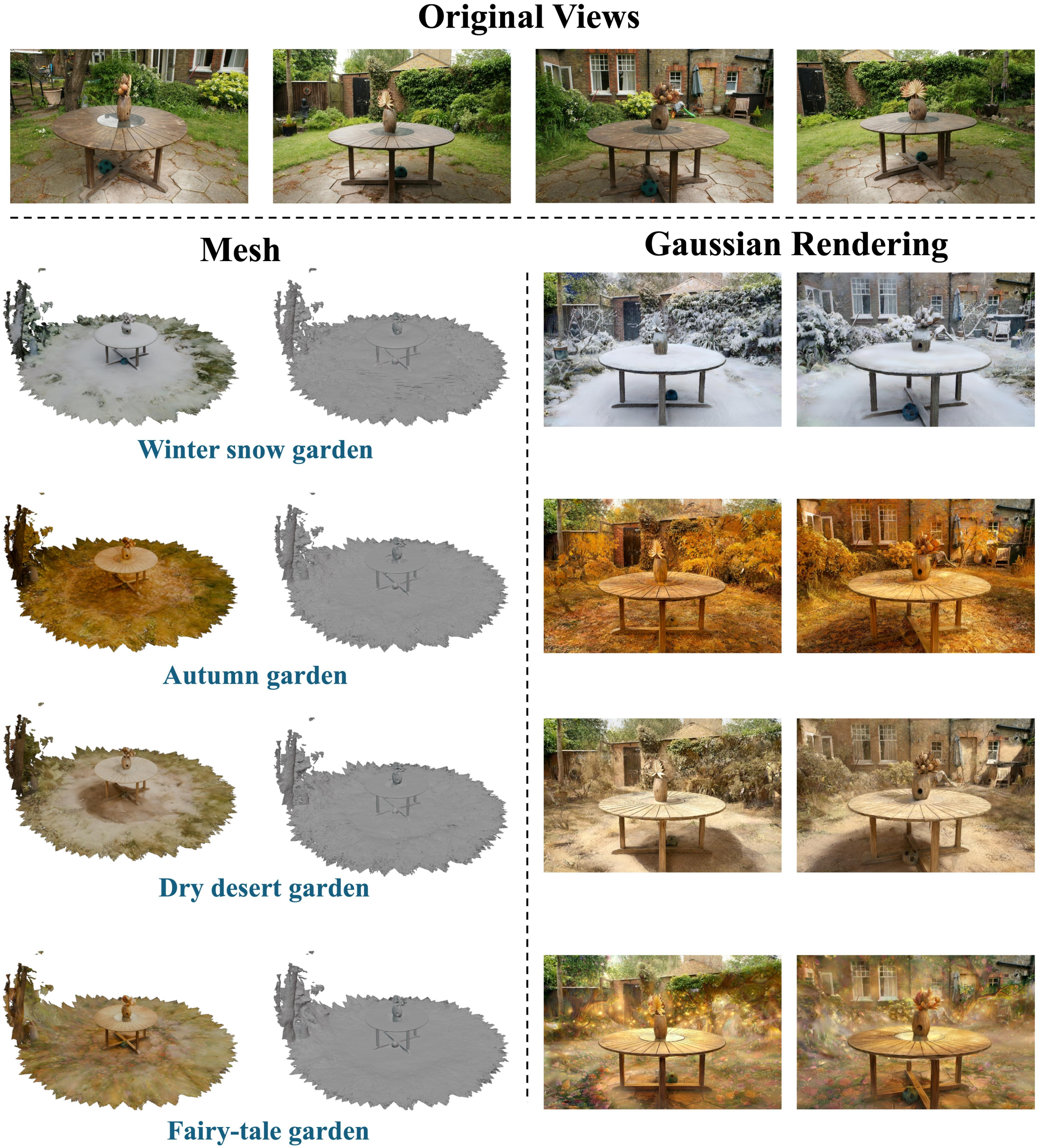}
    \caption{Additional qualitative result of Edit2MeshGS.}
    \label{fig:supp-add-zs-2}
\end{figure*}

\begin{figure*}[p]
    \centering
    \includegraphics[width=\textwidth,height=0.92\textheight,keepaspectratio]{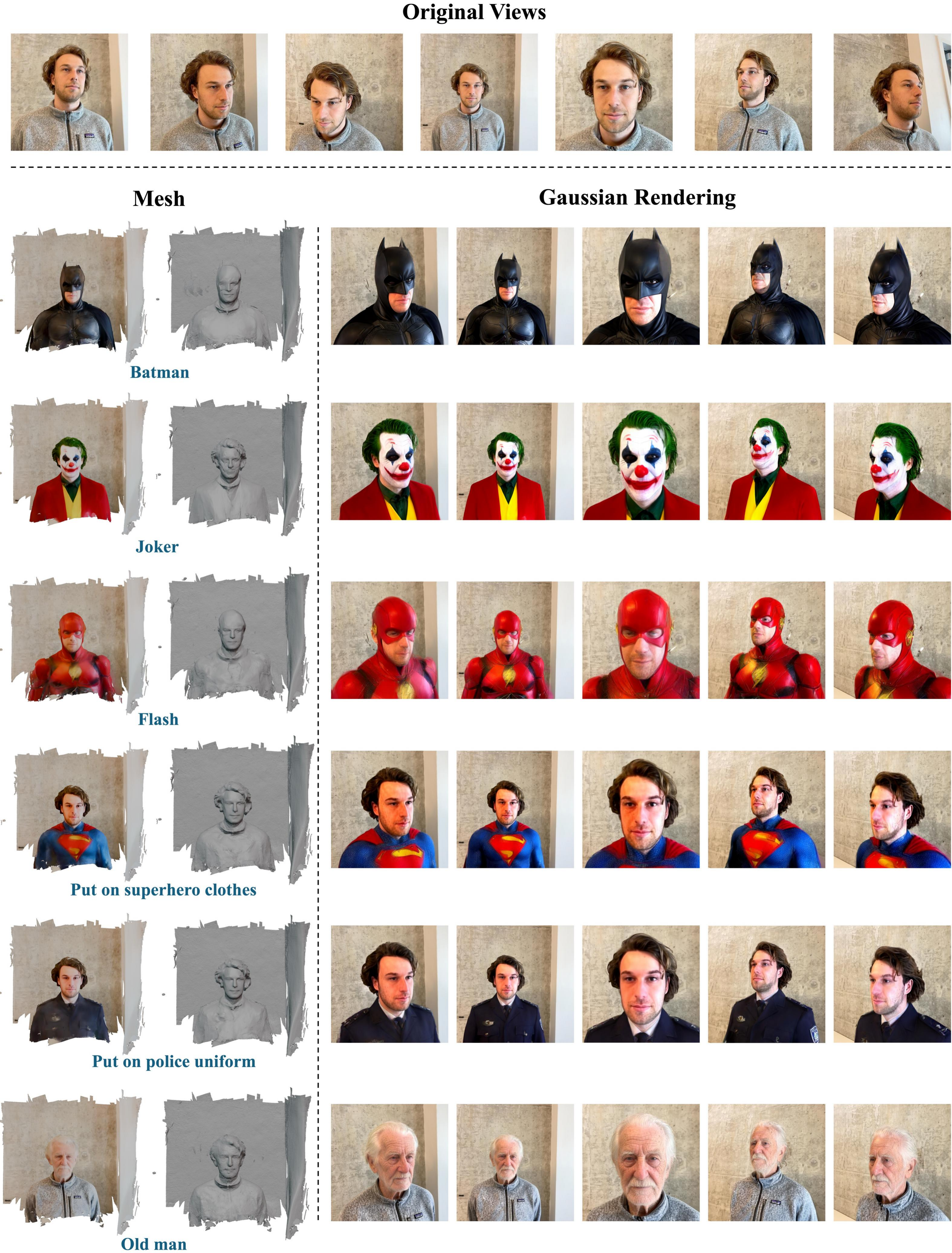}
    \caption{Additional qualitative result of Edit2MeshGS.}
    \label{fig:supp-add-zs-3}
\end{figure*}

\begin{figure*}[p]
    \centering
    \includegraphics[width=\textwidth,height=0.92\textheight,keepaspectratio]{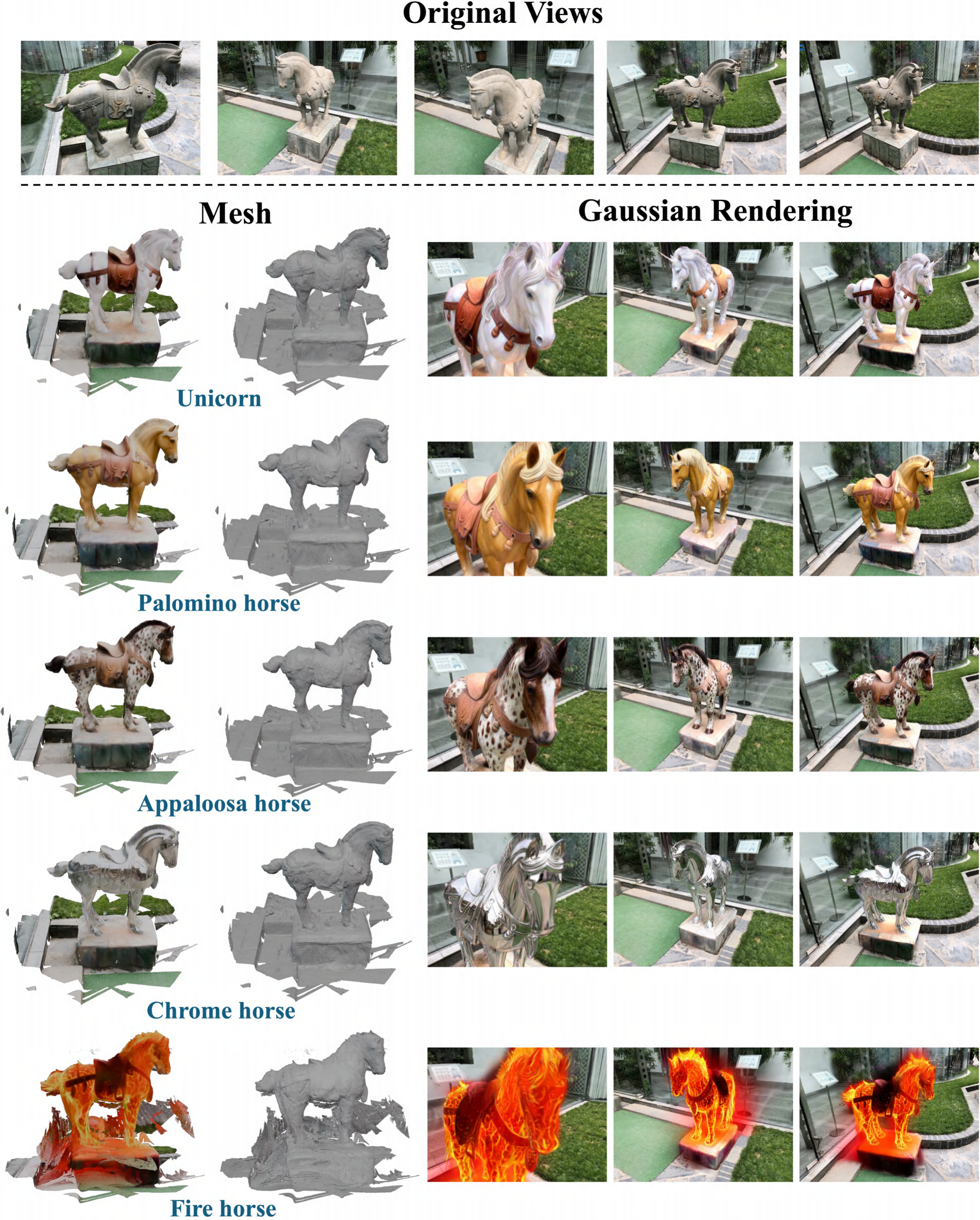}
    \caption{Additional qualitative result of Edit2MeshGS.}
    \label{fig:supp-add-zs-4}
\end{figure*}

Fig.~\ref{fig:flux-scale} provides a qualitative comparison of the 4B and 9B variants; their quantitative results are reported in the main paper. Figs.~\ref{fig:supp-add-zs-1}--\ref{fig:supp-add-zs-4} show additional edited renderings and reconstructed meshes across diverse scenes and instructions. These examples include edits beyond the twelve tasks used for quantitative evaluation.

\section{Editing Instructions and Reference Images}
\label{sec:supp-tasks}

As shown in Fig.~\ref{fig:flux-comp}, using an edited view as the reference can cause FLUX.2 to copy its pose into other views. A separate style reference guides the appearance of the edit while preserving the pose in each input view.

We therefore use a separate style reference image in most tasks to guide appearance consistently across views. Tables~\ref{tab:editing-task-settings}, \ref{tab:editing-task-settings-part2}, and \ref{tab:editing-task-settings-part3} list the editing instructions, semantic prompts, and reference images for the twelve quantitative editing tasks. The Garden tasks use scene-level style edits without source or target semantic prompts.

\clearpage
\begin{table*}[!b]
    \centering
    \begingroup
    \small
    \setlength{\tabcolsep}{1pt}
    \renewcommand{\arraystretch}{1.06}
    \begin{tabular}{@{}K{0.035\textwidth}L{0.065\textwidth}L{0.065\textwidth}L{0.08\textwidth}L{0.61\textwidth}K{0.11\textwidth}@{}}
        \toprule
        ID & Scene & Source & Target & Editing Instruction & Reference \\
        \midrule
        1 & Face & man & spider man & Edit the first image only. Dress the man in a complete Spider-Man costume based on the second reference image, including the red and blue suit, black web pattern, chest emblem, and a full Spider-Man head mask with white eye lenses. The mask must completely cover his entire head, face, hair, ears, skin, and neck, with no human facial features visible. Use the second image only for the costume design and material style. Do not change the original body pose, head tilt, head rotation, gaze direction, face orientation, shoulder angle, camera angle, framing, background, or lighting. Remove all visible hair completely; do not show hair outside, under, or around the mask. & \taskref{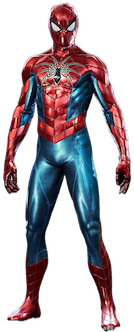} \\
        2 & Face & man & old man & Edit the first image only. Use the second image only as the elderly man appearance reference. Keep the first image man's pose, head angle, gaze direction, face orientation, body orientation, camera angle, framing, background, and lighting unchanged. Do not copy the second image's face identity, pose, body shape, scale, background, or composition. Change only the visible man's appearance into an elderly man, with natural aging details such as white or gray hair, forehead wrinkles, eye wrinkles, subtle skin texture, aged facial features, and a realistic elderly look. Keep the original person's overall facial structure and identity consistent. & \taskref{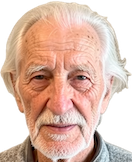} \\
        3 & Stone Bear & stone bear & bear & Edit the first image only. Use the second image only as the firefighter outfit reference. Turn the existing bear statue into a firefighter bear. Keep the first image bear pose, body orientation, head position, head direction, camera angle, framing, base, background, and lighting unchanged. Do not copy the second image's person, pose, body shape, scale, background, or composition. Do not add a new head or move the head. Only change the bear's appearance by adding firefighter gear based on the second image, including a red firefighter helmet, tan firefighter jacket, reflective yellow-gray stripes, black shoulder straps, protective gear, and realistic fabric details. & \taskref{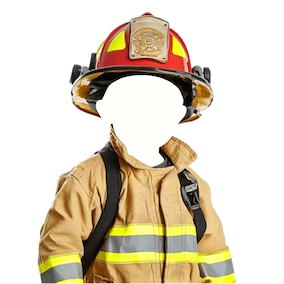} \\
        4 & Stone Bear & stone bear & polar bear & Edit the first image only. Use the second image only as the polar bear appearance reference. Turn the existing bear statue in the first image into a polar bear. Keep the first image bear pose, body orientation, head position, head direction, camera angle, framing, base, background, and lighting unchanged. Do not copy the second image's pose, body shape, scale, background, or composition. Do not add a new head or move the head. Only change the bear's appearance to white polar bear fur based on the second image, with natural fur texture and realistic polar bear coloring. & \taskref{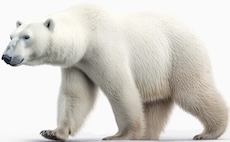} \\
        \bottomrule
    \end{tabular}
    \endgroup
    \caption{Editing instructions and reference images, when used, for the twelve editing tasks. Part 1 includes tasks 1--4.}
    \label{tab:editing-task-settings}
\end{table*}

\clearpage

\begin{table*}[p]
    \centering
    \begingroup
    \small
    \setlength{\tabcolsep}{1pt}
    \renewcommand{\arraystretch}{1.06}
    \begin{tabular}{@{}K{0.035\textwidth}L{0.065\textwidth}L{0.065\textwidth}L{0.08\textwidth}L{0.61\textwidth}K{0.11\textwidth}@{}}
        \toprule
        ID & Scene & Source & Target & Editing Instruction & Reference \\
        \midrule
        5 & Doll & doll & doll & Edit the first image only. Use the second image only as the red hoodie clothing reference. Dress the doll in the first image in a small red hoodie based on the second image. Keep the first image doll pose, face, body shape, fur texture, camera angle, framing, background, and lighting unchanged. Do not copy the second image's shape, scale, background, or composition. Do not change anything except adding the hoodie. & \taskref{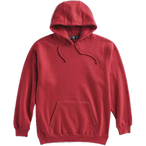} \\
        6 & Doll & doll & robot doll & Edit the first image only. Use the second image only as the robot doll appearance reference. Turn the doll in the first image into a robot doll based on the second image. Keep the first image doll pose, body orientation, head position, face direction, body shape, camera angle, framing, background, and lighting unchanged. Do not copy the second image's pose, body shape, scale, background, or composition. Do not add, move, remove, resize, rotate, or duplicate any body parts. Only change the doll's appearance into a cute robot style, with white metallic panels, blue glowing eyes, mechanical joints, small armor plates, and clean sci-fi details. & \taskref{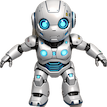} \\
        7 & Garden & -- & -- & Edit the first image only. Use the second image only as the cyberpunk garden patio style reference. Apply a cyberpunk garden patio style to the first image scene. Keep the first image camera angle, framing, layout, table position, centerpiece position, building, trees, plants, ground, and lighting direction unchanged. Do not copy the second image's layout, object positions, camera angle, background, or composition. Do not move, remove, or replace the main objects. Only change the scene style by adding futuristic neon lights, glowing signs, metallic details, wet reflective surfaces, subtle cables, and a dark high-tech atmosphere based on the second image. & \taskref{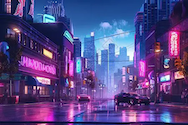} \\
        8 & Garden & -- & -- & Edit the first image only. Use the second image only as the fairy-tale garden style reference. Apply a fairy-tale garden style to the first image scene. Keep the first image camera angle, framing, layout, table position, centerpiece position, building, trees, plants, ground, and lighting direction unchanged. Do not copy the second image's layout, object positions, camera angle, background, or composition. Do not move, remove, or replace the main objects. Only change the scene style by adding a magical storybook atmosphere, lush colorful flowers, soft glowing garden lights, whimsical plants, gentle mist, and warm dreamy colors based on the second image. & \taskref{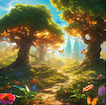} \\
        \bottomrule
    \end{tabular}
    \endgroup
    \caption{Editing instructions and reference images, when used, for the twelve editing tasks. Part 2 includes tasks 5--8.}
    \label{tab:editing-task-settings-part2}
\end{table*}

\clearpage

\begin{table*}[p]
    \centering
    \begingroup
    \small
    \setlength{\tabcolsep}{1pt}
    \renewcommand{\arraystretch}{1.06}
    \begin{tabular}{@{}K{0.035\textwidth}L{0.065\textwidth}L{0.065\textwidth}L{0.08\textwidth}L{0.61\textwidth}K{0.11\textwidth}@{}}
        \toprule
        ID & Scene & Source & Target & Editing Instruction & Reference \\
        \midrule
        9 & Stone Horse & stone horse & unicorn & Edit the first image only. Use the second image only as the unicorn appearance reference. Keep the original horse pose, body orientation, head position, visible body silhouette, saddle position, base position, camera angle, framing, background, and lighting unchanged. Do not copy the second image’s pose, body shape, scale, background, or composition. Do not add wings or any extra body parts; use only the existing visible horse head, face, ears, neck, legs, tail, and body structure from the first image. Do not move, remove, resize, rotate, or duplicate any existing body parts. Change only the visible horse statue surface into a living fantasy unicorn appearance, with a white pearlescent coat, long flowing mane following the original mane area, one elegant spiral horn only on the original forehead, natural horse skin texture, and subtle magical color highlights. & \taskref{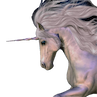} \\
        10 & Stone Horse & stone horse & palomino horse & Edit the image only. Keep the first image horse pose, body orientation, head position, visible body silhouette, saddle position, base position, camera angle, framing, background, and lighting unchanged. Do not add any extra body parts; use only the existing visible horse head, face, ears, neck, legs, tail, and body structure from the image. Do not move, remove, resize, rotate, or duplicate any existing body parts. Change only the visible horse statue surface into a realistic Palomino horse appearance, with a golden coat, light cream mane and tail following the existing mane and tail areas, natural horse skin texture, subtle realistic highlights, and soft color variation. & \taskref{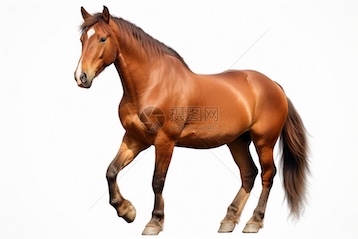} \\
        11 & Bonsai & bonsai flowers & autumn leaves & Turn the bonsai flowers into autumn leaves. & - \\
        12 & Bonsai & bonsai flowers & ice flowers & Change only the bonsai flowers into translucent ice flowers. Keep the bonsai trunk, branches, pot, tablecloth, bicycle, room background, camera angle, framing, and lighting unchanged. Do not change any non-flower objects. & - \\
        \bottomrule
    \end{tabular}
    \endgroup
    \caption{Editing instructions and reference images, when used, for the twelve editing tasks. Part 3 includes tasks 9--12.}
    \label{tab:editing-task-settings-part3}
\end{table*}

\end{document}